\documentclass[fleqn,usenatbib]{mnras}

\usepackage{newtxtext,newtxmath}

\usepackage[T1]{fontenc}

\DeclareRobustCommand{\VAN}[3]{#2}
\let\VANthebibliography\thebibliography
\def\thebibliography{\DeclareRobustCommand{\VAN}[3]{##3}\VANthebibliography}

\usepackage{graphicx}	
\usepackage{amsmath}	

\makeatletter 
\newcount\SOUL@minus
\makeatother  

\title[FRB sub-burst drift law]{Evidence of a shared spectro-temporal law between sources of repeating fast radio bursts}

\author[M. A. Chamma et al.]{Mohammed A. Chamma,$^{1}$\thanks{E-mail: mchamma@uwo.ca} Fereshteh Rajabi,$^{2,3}$
 Christopher M. Wyenberg,$^{1}$ 
\newauthor Abhilash Mathews$^{4}$ and Martin Houde$^{1}$\thanks{E-mail: mhoude2@uwo.ca}
\\
$^{1}$Department of Physics and Astronomy, The University of Western Ontario, 1151 Richmond Street, London, Ontario N6A 3K7, Canada\\
$^{2}$Perimeter Institute for Theoretical Physics, Waterloo, ON N2L 2Y5, Canada\\
$^{3}$Institute for Quantum Computing and Department of Physics and Astronomy, The University of Waterloo, 200 University Ave. West, \\Waterloo, Ontario N2L 3G1, Canada\\
$^{4}$Plasma Science and Fusion Center, Massachusetts Institute of Technology, 77 Massachusetts Avenue, Cambridge, MA 02139, USA\\}

\date{}

\pubyear{2021}

\begin{document}
\label{firstpage}
\pagerange{\pageref{firstpage}--\pageref{lastpage}}
\maketitle
 
\begin{abstract}
We study the spectro-temporal characteristics of two repeating fast radio bursts (FRBs), namely, FRB~20180916B and FRB~20180814A , and combine the results with those from our earlier analysis on FRB~20121102A. The relationship between the frequency drift rate, or slope, of individual sub-bursts and their temporal duration is investigated. We consider a broad sample of possible dispersion measure (DM) values for each source to understand the range of valid sub-burst slope and duration measurements for all bursts and to constrain our results. We find good agreement with an inverse scaling law between the two parameters previously predicted using a simple dynamical relativistic model. The remarkably similar behaviour observed in all sources provides strong evidence that a single and common underlying physical phenomenon is responsible for the emission of signals from these three FRBs, despite their associations with different types of host galaxies at various redshifts. It also opens up the possibility that this sub-burst slope law may be a universal property among repeating FRBs, or indicates a distinct subclass among them.              
\end{abstract}




\section{Introduction}\label{sec:introduction}

Fast radio bursts (FRBs) are short duration ($\sim$ millisecond) bursts of energy at radio wavelengths exhibiting large brightness temperatures ($T_B>10^{32}$~K; \citealt{Lorimer2007,Petroff2019}), suggesting that these signals originate from non-thermal objects through some coherent emission mechanism. Still, the origin and underlying physical mechanism of FRBs remain unknown in spite of the large number of proposed models \citep{Platts2019}. FRB signals also undergo a high level of dispersion as they propagate from the source to the observer, a phenomenon quantified through the dispersion measure (DM). This dispersion results from the wavelength dependence of the refractive index of ionized matter in astronomical media through which radiation travels at varying speeds as a function of frequency. While a first Galactic FRB was recently reported by the CHIME/FRB Collaboration and the STARE2 team toward the Galactic magnetar SGR~1935+2154 \citep{Andersen2020,Bochenek2020}, the DM values measured for most FRBs suggest that these signals must emanate from extragalactic sources. 

Reported FRBs fall into two groups: one-off events and repeaters. While one-off events form the majority of detections, most of our knowledge about FRBs is based on the study of repeaters. At the time of writing, two repeaters (FRB~20121102A and FRB~20180916B, previously known as FRB~121102 and FRB~180916.J0158+65) show periodic behaviours, prompting continued follow-up observations \citep{CHIME2020a,Rajwade2020}. Importantly, the study of dynamic spectra of repeaters reveals interesting patterns. Among these are a downward drift in the central frequency of consecutive sub-bursts with increasing arrival time within an event (the so-called ``sad trombone'' effect), and an average reduction in the temporal duration of individual sub-bursts with increasing frequency \citep{Gajjar2018,Hessels2019,CHIME2019a,CHIME2019c}. 

Several models have been proposed to explain these spectro-temporal characteristics. Some models link these characteristics to the intrinsic radiation mechanism of FRBs \citep{Wang2019,Beloborodov2020b,Metzger2019} or propagation effects (e.g., plasma lensing in \citealt{Cordes2017} or scintillation in \citealt{Simard2020}), while others argue that a combination of both factors can play a part \citep{Hessels2019}. Recently, the detection of the first Galactic FRB \citep{Andersen2020,Bochenek2020} has posed new challenges for existing theoretical models. For example, one sequence of sub-bursts detected toward this source reveals an upward central frequency drift with increasing arrival time (a ``happy trombone'' effect; see Burst 6 of \citealt{Hilmarsson2021}, bursts 24 and 25 in \citealt{CHIME2020a} and FRB~190611 in \citealt{Day2020} for other examples). A few models have anticipated such a possibility for the spectra of FRBs \citep{Simard2020,Rajabi2020b,Beniamini2020}. In particular, \citet{Rajabi2020b} proposed a simple dynamical relativistic model where a descending or an ascending central frequency drift for a sequence of sub-bursts can be explained based on the intrinsic properties of the corresponding FRB source (as discussed in Section \ref{sec:model} and Appendix \ref{sec:sourceparams}). But more importantly, their model also predicts that a steeper frequency drift should be present within individual sub-bursts (henceforth the ``sub-burst slope''\footnote{We denote an FRB event or burst as containing one lone or many pulses of radiation, as is observed for the so-called sad trombone effect, for example. A single pulse in an event containing several pulses will be defined as a sub-burst. \citealt{Rajabi2020b} used the term ``sub-burst drift'' to describe the observed signal delay $t_\mathrm{D}$ as a function of the measured frequency $\nu_\mathrm{obs}$ within an individual sub-burst (as in equation (\ref{eq:drift})). However, since this term is also sometimes found in the literature to denote delays between separate sub-bursts (as in the sad trombone effect), we will instead opt for ``sub-burst slope'' to describe the internal drift within an individual sub-burst to avoid any possible confusion.}) where the slope of the FRB signal as displayed in a waterfall (i.e., the signal intensity as a function of frequency and time) obeys a simple law scaling inversely with the temporal duration of the sub-burst. \citet{Rajabi2020b} further provided evidence for this sub-burst slope behaviour for FRB~20121102A and showed that data taken over a wide range of frequencies for this repeater follow the same law, i.e., the aforementioned inverse scaling of the sub-bursts slope with their corresponding temporal duration. They then argued that this finding implies that the underlying physical process responsible for the signals detected in FRB~20121102A is intrinsically narrow-band in nature, while relativistic motions within the source are required to explain the wide observed bandwidths. 

In this paper, we examine data from two additional repeaters, FRB~20180814A (initially named FRB~180814.J0422+73; \citealt{CHIME2019a}) and FRB~20180916B \citep{CHIME2019b, CHIME2020a}, and show that this form of law is closely shared between these three FRBs originating from host galaxies at different redshifts. We also consider the effect of the chosen DM on the measurements of the sub-burst slope and the temporal duration in order to understand the robustness of this relationship between the sources. Section \ref{sec:model} summarizes the triggered dynamical model that motivates this analysis. Section \ref{sec:analysis} describes the details of the analysis and the handling of the DM for each burst from each source. In Section \ref{sec:results} we highlight the results and discuss the implications of unresolved bursts in our sample, the effect of variations in DM on a measurement of sub-burst slope or drift, and finally explore physical interpretations of our result in the context of the triggered dynamical model. This finding reveals new insights on the underlying physical mechanism at the source of FRB signals and helps refine modelling and characterization efforts.

\subsection{ \texorpdfstring{The triggered dynamical model of \citet{Rajabi2020b}}{} } \label{sec:model}
\citet{Rajabi2020b} introduced a simple dynamical model where a triggering source (e.g., a pulsar or magnetar; see \citealt{Houde2018b}) is located directly behind an FRB source as seen by an observer. An FRB source may contain one or many sub-regions moving towards (or away from) the observer, potentially at relativistic speeds, and from which individual sub-burst signals are emanating. Such a scenario is appropriate for situations where the emitted signal is highly collimated, such as is the case for a radiation process based on Dicke's superradiance which the FRB model proposed by \citealt{Rajabi2020b} is ultimately inspired by \citep{Rajabi2016A,Rajabi2016B,Rajabi2017,Mathews2017,Houde2018a,Houde2018b,Rajabi2019,Rajabi2020}. Under such conditions it can be shown that the slope of a single sub-burst signal (for repeaters an event can contain several sub-bursts) obeys the following relation
\begin{align}
    \frac{1}{\nu_\mathrm{obs}}\frac{d\nu_\mathrm{obs}}{dt_\mathrm{D}} = -\frac{A}{t_\mathrm{w}},\label{eq:drift}
\end{align}
\noindent where $\nu_\mathrm{obs}$, $t_\mathrm{w}$ and $t_\mathrm{D}$ are the frequency, the temporal duration of the sub-burst and the delay before its appearance (in relation to the arrival of the trigger) as measured by the observer. The systemic parameter $A\equiv \tau_\mathrm{w}^\prime/\tau_\mathrm{D}^\prime$ with $\tau_\mathrm{w}^\prime$ and $\tau_\mathrm{D}^\prime$ the corresponding sub-burst proper temporal duration and delay in the reference frame of the corresponding FRB sub-region, respectively.

Following the model of \citet{Rajabi2020b}, the temporal duration of an FRB sub-burst in the observer rest frame is given by
\begin{align}
    t_\mathrm{w} = \tau_\mathrm{w}^\prime\frac{\nu_0}{\nu_\mathrm{obs}},\label{eq:tw_SM}
\end{align}
\noindent where $\nu_0$ is frequency of emission in the FRB rest frame. Equation (\ref{eq:tw_SM}) clearly predicts an inverse relationship between the observed FRB temporal width and frequency, which had previously been noticed and studied. For example, a verification of this effect can be found in Figure 7(b) of \citet{Gajjar2018} for the case of FRB~20121102A\footnote{As noted in \citet{Rajabi2020b}, the temporal narrowing effect with observed frequency is likely more pronounced than seen in Figure 7(b) of \citet{Gajjar2018} as the authors do not measure the duration of individual sub-bursts $t_\mathrm{w}$ but rather that of the whole event. The true values for $t_\mathrm{w}$ are therefore likely to be lower and the sub-burst slope more pronounced in several instances.}. Although the measurements of burst temporal duration  exhibit a fair amount of scatter at a given frequency (which could also be inherent to $\tau_\mathrm{w}^\prime$ in equation (\ref{eq:tw_SM})), the predicted behaviour is consistent with the observations. 

\citet{Rajabi2020b} also derived the following equation for the relative drift in the observed central frequency of a sequence of sub-bursts with increasing arrival time 
\begin{align}
    \frac{\Delta\nu_\mathrm{obs}}{\Delta t_\mathrm{D}} = \frac{\nu_\mathrm{obs}}{\nu_0}\frac{d\nu_\mathrm{obs}}{d\tau_\mathrm{D}^\prime},\label{eq:sad_trombone_SM}
\end{align}
\noindent where the term on the left-hand side is for the relative central frequency drift and $\tau_\mathrm{D}^\prime$ is the proper temporal delay between the arrival of the trigger and the emission of the ensuing sub-burst in the FRB rest frame. The derivative $d\nu_\mathrm{obs}/d\tau_\mathrm{D}^\prime$ is a physical parameter characterizing the environment of the FRB source, which determines whether the sequence of sub-bursts has the appearance of a ``sad'' ($d\nu_\mathrm{obs}/d\tau_\mathrm{D}^\prime<0$) or ``happy trombone'' ($d\nu_\mathrm{obs}/d\tau_\mathrm{D}^\prime>0$; see \citealt{Rajabi2020b} for more details). Equation (\ref{eq:sad_trombone_SM}) predicts that the central frequency drift should scale linearly with $\nu_\mathrm{obs}$, which has previously been verified for FRB~20121102A over a wide range of frequencies. This can be asserted, for example, from Figure 3 (top panels) of \citet{Hessels2019}, Figure 6 of \citet{CHIME2019c} and Figure 4 of \citet{Caleb2020}. Extended Figure 9 of \citet{PastorMarazuela2020} shows the trend for FRB~20180916B. This observed dependency could not be realized if $\nu_0$ changed significantly in equation (\ref{eq:sad_trombone_SM}). For example, a change of 50\% in $\nu_0$ would markedly affect the appearance of the figures.

We also note that within the context of our triggered model, individual sub-bursts belonging to a single FRB event all results from the same background trigger signal. Their sequence of appearance in time, as seen by the observer, will vary depending on the physical properties of the medium where individual sub-bursts emanate from (which will affect the delay time in the corresponding rest frame $\tau_\mathrm{D}^\prime$) and its velocity $\beta$ (and therefore the frequency $\nu_\mathrm{obs}$ relative to the observer). Although we expect sub-bursts belonging to a single FRB event to be clustered in time, it is also possible that sub-bursts belonging to different events be observed relatively closely in time.

\section{\texorpdfstring{Burst Analysis}{}}\label{sec:analysis}
Although equation (\ref{eq:drift}) was tested and verified for FRB~20121102A in \citet{Rajabi2020b} using previously published data covering more than a decade in frequency \citep{Michilli2018,Gajjar2018,CHIME2019c}, it was not known at the time whether it applies equally well to other repeating FRBs. We therefore retrieved and analyzed previously published data for two other sources discovered by the CHIME/FRB Collaboration \citep{CHIME2020}, namely FRB~20180916B \citep{CHIME2020a} and FRB~20180814A \citep{CHIME2019a}. These data are all contained within the CHIME/FRB spectral band (approximately 400--800~MHz) and the corresponding dynamic spectra were analyzed using the two-dimensional autocorrelation technique introduced in \citet{Hessels2019}, resulting in estimates for the sub-burst slope ($d\nu_\mathrm{obs}/dt_\mathrm{D}$) and temporal duration ($t_\mathrm{w}$). See Appendix \ref{app:autocorr} for more details. These data sources were chosen purely due to their ease of accessibility and the support available. Ultimately we aim to extend this analysis to as many sources and bursts as possible.

\subsection{\texorpdfstring{The effect of the Dispersion Measure (DM)}{}}\label{sec:effectofdm}
Since the measurement of any drift rate (or almost any other spectro-temporal feature) is strongly dependant on the DM that is used to dedisperse a waterfall, and since the DM of a source can potentially vary from burst to burst as well as with time, we studied the variation of our slope and temporal duration measurements for each sub-burst at different choices of DM. Dedispersion can be performed by optimizing either the signal-to-noise (S/N) or a structure parameter and can result in different values found for the DM depending on the burst (e.g. Fig 1 of \citealt{Gajjar2018}). In particular, an algorithm seeking to choose a DM by maximizing S/N might superimpose the individual sub-bursts of a complex FRB event and yield a DM value that is higher than a structure optimizing algorithm. For bursts with components that are not clearly resolved it becomes ambiguous which algorithm is most accurate and the precision in the DMs determined burst to burst can be much narrower than the variations in the DM observed overall for a source \citep{CHIME2020a}. It therefore becomes difficult to uncouple FRB characteristics from the nature of the medium in order to study relationships between spectro-temporal features as we hope to do. One option is to use the DM found on a burst by burst basis. However, doing this can become a complicated process of verifying that the DM algorithm choice is appropriate, which will often be ambiguous for smeared bursts where it is not clear if it consists of multiple components or not. Without a detailed understanding of the emission mechanism, the medium, the source, and the resulting DM distribution as a function of time, it is in fact much simpler and more conservative to choose a DM range as wide as possible based on the history of DMs found for the source. We shall see that despite the significant uncertainties this choice entails, the data still point to the existence of an inverse trend between the sub-burst slope and temporal duration for the three sources considered here.

\begin{table}
    \resizebox{\columnwidth}{!}{
    \begin{centering}
    \begin{tabular}{lll}
    \hline
    \textbf{Source} & \textbf{Data source} & \textbf{DM Range} (pc/cm$^{3}$)\tabularnewline
    \hline
    FRB~20121102A & \citealt{Michilli2018} & 554.1--565.3\tabularnewline
    \hline
    FRB~20121102A & \citealt{Gajjar2018} & 555--570 (555--583) \tabularnewline
    \hline
    FRB~20180916B & \citealt{CHIME2020a} & 346.82--349.82 \tabularnewline
    \hline
    FRB~20180814A & \citealt{CHIME2019a} & 188.7--190.0 \tabularnewline
    \hline
    \end{tabular}
    \end{centering}
    }
\caption{The range of DMs used to determine the range of possible values of each sub-burst slope and duration. These are chosen to be as wide as possible while still obtaining reasonable sub-burst slope measurements. In general, the published history of DMs found for a source (all bursts considered) determines the range used, with some DMs on the higher end excluded due to resulting positive sub-burst slopes or distortion. The DM range in parentheses is used specially for Burst 11D from FRB~20121102A in \citealt{Gajjar2018} due to its high S/N optimized DM. See the text for more details.}
\label{tab:dmranges}
\end{table}

Table \ref{tab:dmranges} shows the DM ranges chosen for each source and dataset. We aim to consider as broad a range of DMs as possible while still obtaining reasonable sub-burst slope measurements. For the data used from \citealt{Michilli2018} DM variations are estimated by those authors to be $\lesssim$ 1\% of 559.7 pc/cm$^{3}$, and we therefore consider a range of 554.1--565.3 pc/cm$^{3}$. For the data from \citealt{Gajjar2018}, due to availability, we use the sub-bursts in Burst 11A and Burst 11D. A structure optimized DM for 11A is found at 565 pc/cm$^{3}$, and their Figure 1 indicates that DMs between 555-570 are also close to optimal, so we adopt this range. For Burst 11D, due to a lack of structure we consider a range of 555-583 to be closer to its S/N optimized DM, however higher DMs are excluded as the sub-burst slopes start to become positive (which are not physical according to our model and in general usually indicate overdedispersion, as described in Section \ref{sec:exclusions}). For data from \citealt{CHIME2020a} on FRB~20180916B a precise DM of 348.82 $\pm$ 0.05 pc/cm$^{3}$ is found for one of the bursts, but burst-to-burst the DM can range from 348.7-350.2. We therefore choose a mid-point of about 348.82 pc/cm$^{3}$ and adopt a range of 346.82-349.82 pc/cm$^{3}$. The lower value for the start of the range is chosen to push the limit of acceptable DMs while still obtaining reasonable sub-burst slope measurements. We stay away from the higher end of the observed range due to the sub-burst distortion and positive slopes observed for most cases at that high of a DM. Finally, for data from \citealt{CHIME2019a} on FRB~20180814A, due to the structure present in the bursts, we extend the full range of structure optimized DMs found (188.9-190 pc/cm$^{3}$) to 188.7-190 pc/cm$^{3}$. We ignore the higher S/N optimized DMs due to the component overlap and distortion observed when dedispersing to those DMs.

For each source, we generate a grid of DMs over the range chosen and dedisperse all bursts to each DM before performing an autocorrelation analysis. The grid spacing varies from $\Delta$DM $\simeq$ 0.1--2~pc/cm$^3$ depending on the source, yielding approximately 10--20 trial DMs in each case. For FRB events with multiple components like Burst 11A from \citealt{Gajjar2018} for FRB~20121102A, the components are separated manually by finding valleys in the corresponding time series of the data. When necessary these components are padded with a background sample of the waterfall so that there is a wide enough temporal extent to properly dedisperse the burst. Some bursts are not clearly resolved, but wherever there is indication that the slope suddenly changes mid-burst a manual attempt is made to separate the components. 

The autocorrelation analysis (see Appendix \ref{app:autocorr}) is then performed for all dedispersed waterfalls to obtain sub-burst slope and temporal duration measurements for every burst at each DM \citep{Hessels2019}. We use these data to determine the range of possible values for each measurement. Examples of a waterfall for every sub-burst used in this analysis with their corresponding autocorrelation are shown in Figures \ref{fig:frb121102card} -- \ref{fig:frb180814card_2} at the end of the paper, displayed for one of the trial DMs. The range found for each of these measurements is much larger than the parameter uncertainty resulting from the underlying two-dimensional Gaussian fit of the autocorrelation function used to evaluate them. Since the true underlying DM distribution for each source appears to be narrower than the DM range we have used (considering the distribution so far implied by published DMs and knowing that the distribution can change with time), the range of values found this way must be larger than the range implied by the true uncertainties for each measurement. We therefore treat the range of values found by this analysis as upper-limit estimates of the real measurement uncertainties.

\subsection{\texorpdfstring{Measurement exclusions and fitting}{}}\label{sec:exclusions}
With the measurements for each sub-burst at all trial DMs found, there remain measurements that are unconstrained and/or non-physical that need to be discarded before fitting. As previously mentioned we discard any positive sub-burst slope measurements which are non-physical under our model, as well as measurements where the value and/or uncertainties approaches infinity, as is the case for sub-bursts that become near vertical or circular in their autocorrelation.

The result of this exclusion process is that out of a total of 41 sub-bursts analysed, we retain all the measurements made for 28 sub-bursts over the entirety of the DM ranges specified in Table \ref{tab:dmranges}. For the remaining 13 sub-bursts the measurements excluded were taken at the higher end of the DM range, since they yielded positive sub-burst slopes. For all but one of the sub-bursts treated this way the DM range is slightly further restricted. The exception being one burst from FRB~20180814A, where the DM range is limited from 188.7 -- 190.0 pc/cm$^{3}$ to 188.8 -- 188.9 pc/cm$^{3}$. We specifically identify (i.e., circle) these sub-bursts when displayed in  Figures \ref{fig:sub-burst_drift} and \ref{fig:sub-burst_drift_SM} below to indicate the smaller range of DMs used.

Using this set of sub-burst slope and temporal duration measurements we find a fit to equation (\ref{eq:drift}) at each DM and compute the reduced-$\chi^2$ to select a representative DM for each source (i.e., the DM with the reduced-$\chi^2$ closest to unity). We do not perform a fit at the highest DM of 583 pc/cm$^3$ for FRB20121102A as Burst 11D is the only point with a valid measurement at that DM. The representative DMs found are 558.8 and 568.3~pc/cm$^3$ for FRB~20121102A for the data from \citealt{Michilli2018} and \citealt{Gajjar2018}, respectively, 348.82~pc/cm$^3$ for FRB~20180916B and 188.8~pc/cm$^3$ for FRB~20180814A.

\section{Results and Discussion}\label{sec:results}
\begin{center}
    \begin{figure*}
        \centering
        \includegraphics[trim=0 0 0 0, clip,width=1.\textwidth]{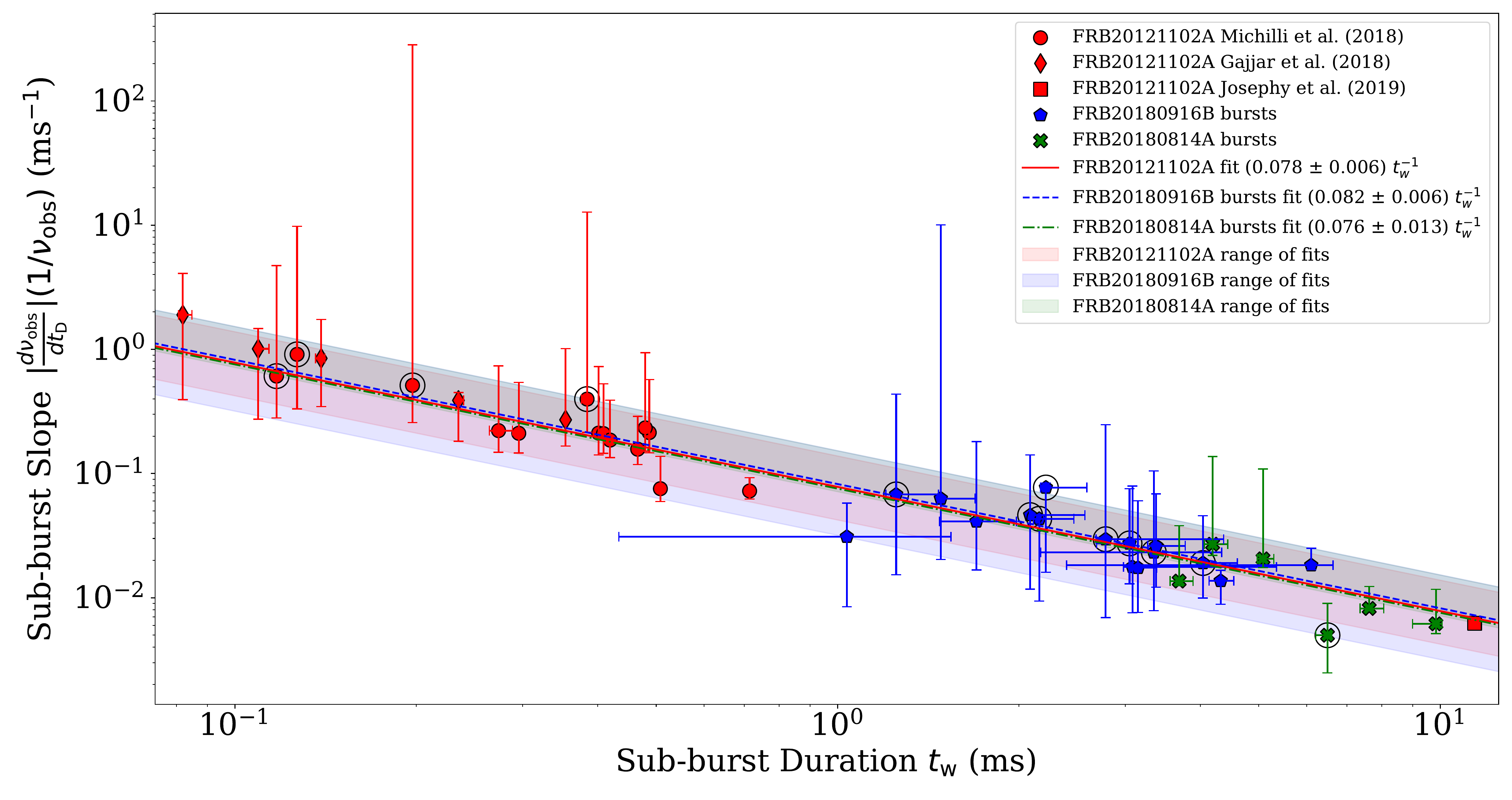}
        \caption{A plot of $\left|d\nu_\mathrm{obs}/dt_\mathrm{D}\right|\left(1/\nu_\mathrm{obs}\right)$ vs. $t_\mathrm{w}$ for bursts from FRB~20121102A (red circles, diamonds and square; \citealt{Gajjar2018,Michilli2018,CHIME2019c}), FRB~20180916B (blue pentagons; \citealt{CHIME2020a}) and FRB~20180814A (green crosses; \citealt{CHIME2019a}). The sub-burst slope $d\nu_\mathrm{obs}/dt_\mathrm{D}$ and duration $t_\mathrm{w}$ were obtained using the two-dimensional autocorrelation technique of \citet{Hessels2019}, while the center frequency $\nu_\mathrm{obs}$ was estimated from the corresponding dynamic spectra. Each burst was dedispersed to a grid of trial DMs over the range specified by Table \ref{tab:dmranges} and the measurements were repeated. The one point from \citealt{CHIME2019c} was not part of the same analysis and is shown for reference. The red, blue and green lines are for fits of the function $A/t_\mathrm{w}$ on the FRB~20121102A, FRB~20180916B and FRB~20180814A data, respectively, at the DM within the range of trial DMs for which the reduced-$\chi^2$ of the fit was closest to unity, and are difficult to distinguish from one another. All points for a given source (except for the \citealt{CHIME2019c} datum) are of measurements made at the same DM used for the corresponding fit. The capped lines at each point represent the range of possible measurements obtained via the autocorrelation analysis for different DMs over the DM ranges chosen. As discussed in Section \ref{sec:effectofdm}, these are used in lieu of, and are larger than, the difficult to determine true measurement uncertainties. The circled points indicate sub-bursts that required a limited DM range to constrain their measurements (see Section \ref{sec:exclusions}). The shaded regions represent the range of fits found when using measurements obtained at other DMs in the range. These regions overlap significantly, but indicate the possibility of unique and distinct fits between the three sources within the range of possible DMs chosen.} 
        \label{fig:sub-burst_drift}
    \end{figure*}
\end{center}

We show in Figure \ref{fig:sub-burst_drift} the results of our analysis, where the sub-burst slope (normalized to the frequency of observation $\nu_\mathrm{obs}$) is plotted against the temporal width $t_\mathrm{w}$ for the three FRBs. Normalizing the sub-burst slope has the advantage of allowing us to combine the different sources on the same graph irrespective of the frequency of observation, shifts due to the dynamical Doppler effect or cosmological redshift. Furthermore, we note that equation (\ref{eq:drift}) is also insensitive to temporal scaling transformations. For example, interstellar scintillation, which brings a temporal broadening scaling inversely with the fourth power of the frequency, will have no effect on our analysis. The only consequence being a shift of data points along the specific law characterized by the parameter $A$ in equation (\ref{eq:drift}). The points displayed in Figure \ref{fig:sub-burst_drift} are the measurements of each sub-burst obtained at the representative DM described at the end of Section \ref{sec:exclusions}, and the capped lines represent the range of possible measurements over the DM range considered. 

Examination of Figure \ref{fig:sub-burst_drift} reveals that the inverse relationship between the two parameters is clearly seen for all sources on the graph for values ranging over two orders of magnitude for both the normalized sub-burst slope and the temporal duration. Also shown in the figure are fits for the predicted function $A/t_\mathrm{w}$ (see equation (\ref{eq:drift})) for the three sources at their representative DMs (see end of Section \ref{sec:exclusions}), with $A = 0.078\pm0.006$, $0.082\pm0.006$ and $0.076\pm0.013$ for FRB~20121102A, FRB~20180916B and FRB~20180814A, respectively. The shaded regions for each source represent the range of fits found when considering all DMs in the adopted range. The corresponding range of fit parameters are found to be $A = 0.042 - 0.138$ for FRB~20121102A, $A = 0.032 - 0.153$ for FRB~20180916B and $A = 0.071 - 0.152$ for FRB~20180814A. These regions overlap significantly, but leave open the possibility of unique and distinct fits between the three sources. 

\begin{figure*}
\centering
\includegraphics[width=0.95\textwidth]{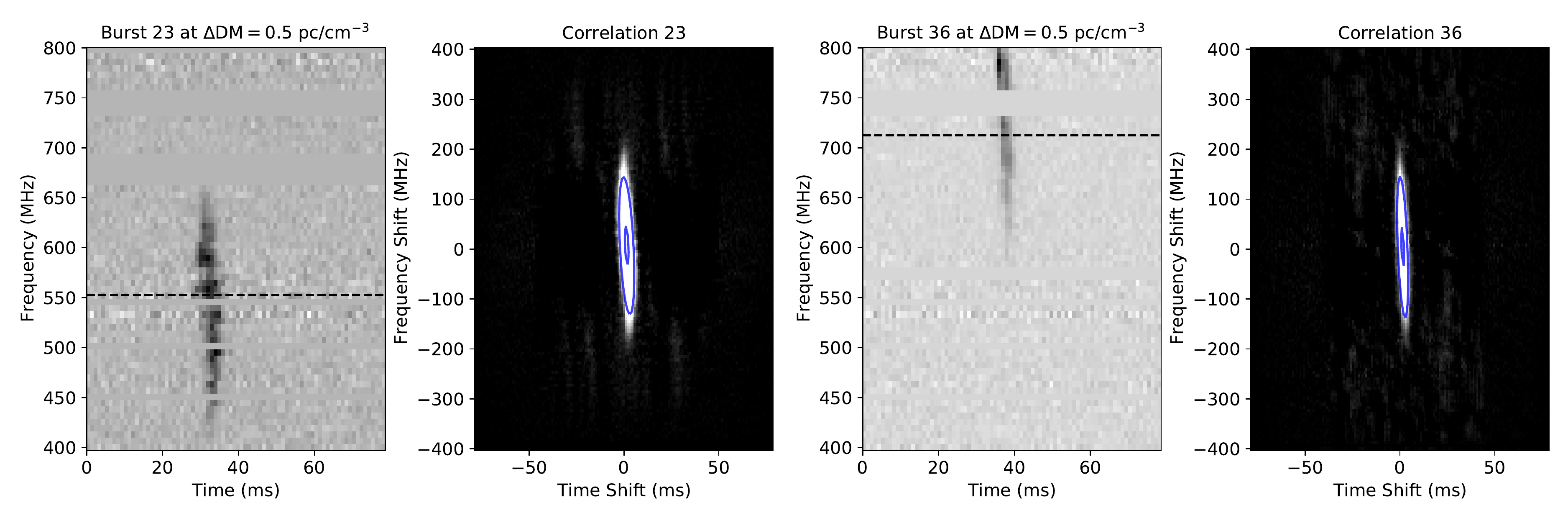}
\includegraphics[width=0.95\textwidth]{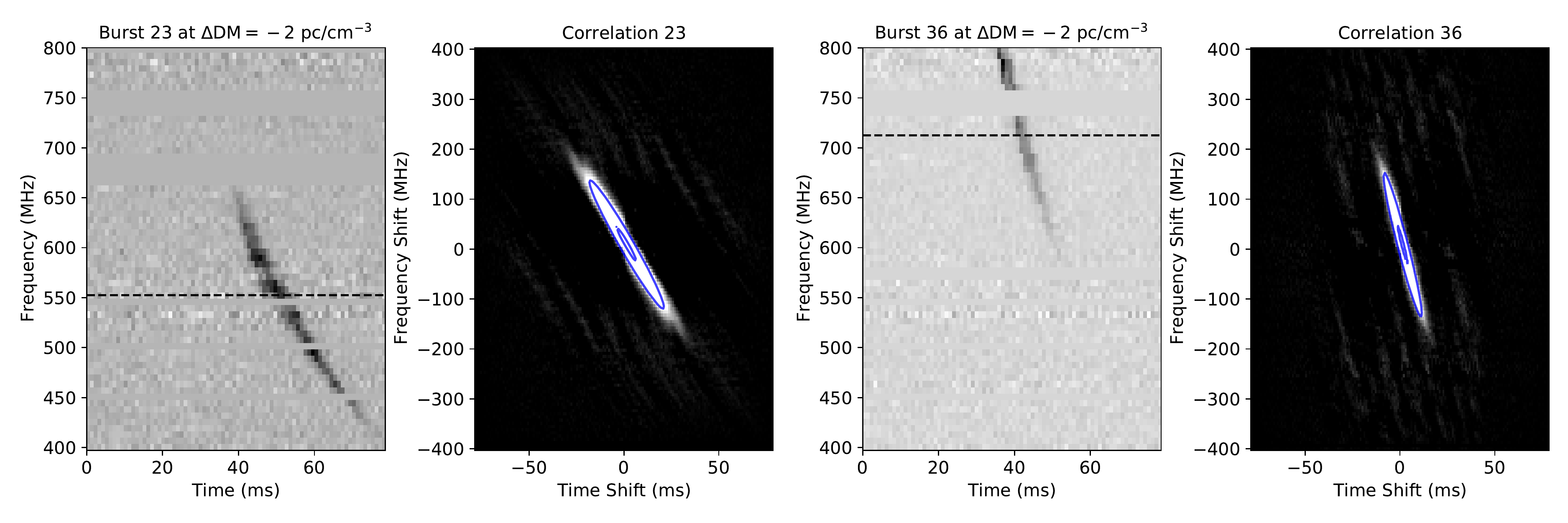}
\caption{Changes to waterfalls and autocorrelations due to variations in the DM. Sub-bursts 23 (first column) and 36 (third column) for FRB~20180916B are shown with their autocorrelation functions (second and fourth columns, respectively) for two offsets $\Delta\mathrm{DM} = 0.5\;\mathrm{pc\,cm}^{-3}$ (top row) and $-2\;\mathrm{pc\,cm}^{-3}$ (bottom row) from the representative value chosen for our analysis (i.e., $\mathrm{DM} = 348.82\;\mathrm{pc\,cm}^{-3}$). The rotations brought about by the small changes in DM are clearly seen in both the waterfall and autocorrelation functions. See Figure \ref{fig:frb180916card} for these bursts  dedispersed to their representative DM.}\label{fig:rotation_stamp}
\end{figure*}

A few important consequences are to be noted from the results presented in Figure \ref{fig:sub-burst_drift}. 
First, and most importantly, the inverse relationship between the sub-burst slope and temporal duration is observed to be independent of the selection of the DM. That is, there is an unmistakeable tendency for narrower sub-bursts at higher slopes, and vice-versa, irrespective of the DM chosen. Even with the ranges of measurements seen, Figure \ref{fig:sub-burst_drift} strongly suggests each source obeys a form of equation (\ref{eq:drift}), where we see a clear decrease in the magnitude of slopes with increasing sub-burst durations.

Second, we note the possibility that not only is the inverse relationship between the sub-burst slope and temporal duration verified for the three sources, but they do so with similar values for $A$ in equation (\ref{eq:drift}) at their representative DMs. The different fits to this systemic parameter are similar given their uncertainties, and it is difficult to visually distinguish between the corresponding curves. This closeness between the values obtained for $A$ is rather remarkable and suggests the existence of a single and common underlying physical phenomenon responsible for the emission of FRB signals in the three sources. This is significant because these FRBs are associated with different types of host galaxies at various redshifts. More precisely, FRB~20121102A is localized to a low-metallicity irregular dwarf galaxy at a redshift $z = 0.193$ \citep{Tendulkar2017}, while the redshift of FRB~20180814A is estimated to be $z\leq0.1$ \citep{CHIME2019a}. Furthermore, the candidates for the host galaxy of FRB~20180814A are not consistent with those harboring long gamma-ray bursts (LGRBs) or superluminous supernovae (SLSNe), unlike the host galaxy of FRB~20121102A \citep{Li2019}. As for FRB~20180916B, it is precisely localized to a star-forming region in a massive spiral galaxy at a redshift $z=0.0337$ \citep{Marcote2020}. This source is the closest known extragalactic FRB, whose host galaxy does not show signatures of a strong magnetic field nor a radio counterpart as reported for FRB~20121102A. The similarities in the values for $A$ between the three sources also suggests that the sub-burst slope law can become a suitable method for making small simplifying adjustments to the DM of waterfalls of repeating FRBs, once the dominant dispersion effects due to the interstellar and intergalactic media are accounted for. The resulting choice of DM would have the advantage of being rooted on a simple physical model resting on the relativistic nature of FRBs, and can be used to simplify analyses with large sample sizes by avoiding the complexity that can arise when choosing a DM based on the S/N or other structure criteria.

As was discussed in \citet{Rajabi2020b}, the three predictions made by their simple dynamical model (i.e., the narrowing of sub-bursts width $t_\mathrm{w}$ with increasing frequency $\nu_\mathrm{obs}$, the sad or happy trombone effect and the sub-burst slope law discussed here) provide strong evidence that the underlying physical phenomenon is narrow-band in nature. This is because the dependencies on $\nu_\mathrm{obs}$ and the frequency of emission in the FRB rest frames $\nu_0$ for the three predicted relationships are such that it would be difficult to envision how they could be realized through the data if $\nu_0$ was allowed to vary substantially (see Section \ref{sec:narrownature} for more details). Although data over a significant range of observed frequency is currently only available for FRB~20121102A (and constitutes the basis of the analysis presented in \citealt{Rajabi2020b}), the fact that FRB~20180916B and FRB~20180814A follow the same law renders it reasonable to expect that the conclusions reached for FRB~20121102A also apply to them.     

We can use this information with our model to further characterise the environment of the sources responsible for the detected bursts. Indeed using the extensive data available for FRB~20121102A one can estimate, although with limited precision at this point, the maximum Lorentz factor and the rest frame frequency of emission $\nu_0$. To do so we will assume highly simplified conditions, i.e., that the different FRB reference frames from which the individual sub-bursts emanate either move towards or away from the observers with the same range of speeds. We will denote by $\beta^+>0$ and $\beta^-=-\beta^+$ the maximum velocities (divided by the speed of light) towards and away from the observer, respectively, with corresponding observed frequencies $\nu_\mathrm{obs}^\pm$. It is then straightforward to show that, under this assumption,
\begin{align}
    \beta^+ & = \frac{\nu_\mathrm{obs}^+ - \nu_\mathrm{obs}^-}{\nu_\mathrm{obs}^+ + \nu_\mathrm{obs}^-}\label{eq:beta+} \\
    \nu_0^2 & = \nu_\mathrm{obs}^+\nu_\mathrm{obs}^-.\label{eq:nu0}
\end{align}
\noindent Using $\nu_\mathrm{obs}^+\simeq7.5$~GHz and $\nu_\mathrm{obs}^-\simeq630$~MHz we find $\beta^+\approx0.9$ and $\nu_0\approx2.6$~GHz for FRB~20121102A (taking into account its known redshift $z=0.193$ from \citealt{Tendulkar2017}; see Section \ref{sec:narrownature} for more details). Evidently, the accuracy for these estimates is set and limited by the frequency coverage of the existing data and is likely to change as more detections are acquired. For example, confirming the purported detection of signals at $111$~MHz from \citet{Fedorova2019} would further increase $\beta^+$ and bring down $\nu_0$ on the order of 1~GHz. At any rate, these results imply that FRB~20121102A is potentially very strongly relativistic.  

We also know that the spectral width $\Delta\nu_\mathrm{obs}$ associated to individual sub-bursts for FRB~20121102A scales as $\Delta\nu_\mathrm{obs}\sim0.16~\nu_\mathrm{obs}$ on average (see Figure 6 in \citealt{Rajabi2020b} or Figure 5 in \citealt{Houde2018b}). This spectral extent is the result of motions (through the Doppler effect) within a given FRB rest frame from where a sub-burst centred at $\nu_\mathrm{obs}$ originates. As discussed in Appendix \ref{sec:sourceparams}, the observed spectral width is constrained through
\begin{align}
    2\Delta\beta^\prime \leq \frac{\Delta\nu_\mathrm{obs}}{\nu_\mathrm{obs}} \leq \frac{2\Delta\beta^\prime}{1-{\Delta\beta^\prime}^2},\label{eq:spectral_width}
\end{align} 
\noindent where the motions in the FRB rest frame are contained within $\pm\Delta\beta^\prime$. We thus find $\Delta\beta^\prime\sim0.08$ with equation (\ref{eq:spectral_width}) for this source. 

We thus have a picture where FRB~20121102A and similar sources would consist of systems within which a number of spatially distinct FRB rest frames, whose motions cover a wide range of velocities (some highly relativistic relative to the observer; $\left|\beta\right|\lesssim 0.9$ for FRB~20121102A), are responsible for the emission of individual sub-bursts. In turn, each such rest frame is also host to mildly relativistic motions ($\left|\Delta\beta^\prime\right|\lesssim 0.08$ for FRB~20121102A), which are responsible for the observed wide spectral widths of sub-bursts.

In the following sections we discuss the the ambiguity between slope and drift rate measurements that arises when sub-bursts are unresolved, the effect of DM variations on the autocorrelation of waterfalls, as well as the effects of noise and missing data on the measurement of the sub-burst slope and temporal duration. We also further discuss the determination of physical parameters of the source. 

\begin{center}
\begin{figure}\hspace*{-0.5cm}
\includegraphics[width=1\columnwidth]{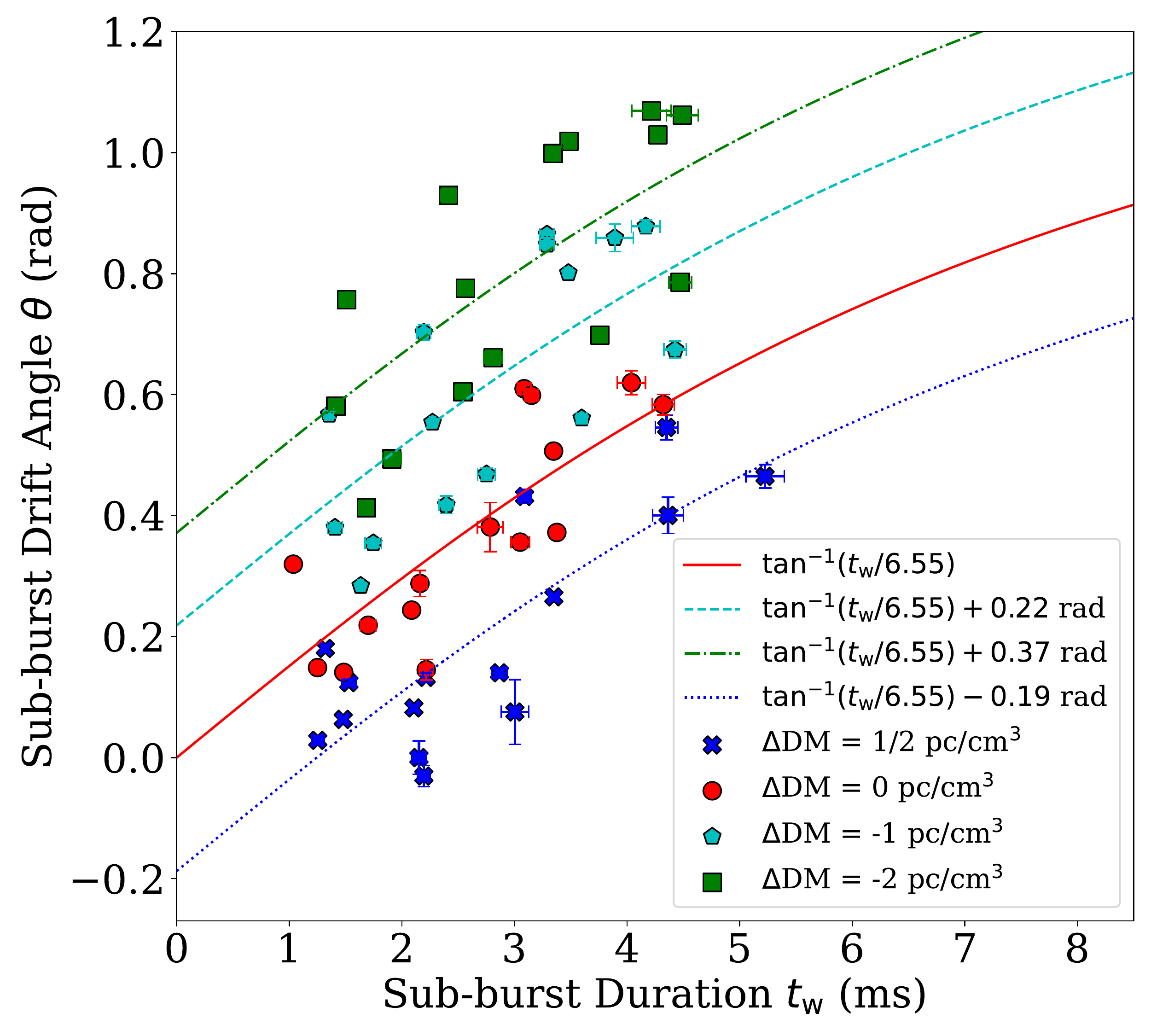}
\caption{The fit angle $\theta$ vs. temporal duration $t_{\mathrm{w}}$ from sub-bursts dedispersed to small variations in the DM for the source FRB~20180916B. Red circles are sub-bursts at $\Delta\mathrm{DM}=0$, which corresponds to a $\mathrm{DM} = 348.82\;\mathrm{pc\,cm}^{-3}$. Blue crosses, cyan pentagons, and green squares are sub-bursts dedispersed to $\Delta\mathrm{DM}=0.5$, $-1$, and $-2\;\mathrm{pc\,cm}^{-3}$, respectively. Error bars indicate the parameter fitting uncertainty. The red curve is the fit to the red circles and is of the form given in equation (\ref{eq:theta_SM}), derived from the dynamical model described in the main text. Blue, cyan, and green curves are obtained by adding a rotation (i.e., adding an angle) to the $\Delta\mathrm{DM}=0$ model. As discussed in Section \ref{sec:dmrotates} this plot demonstrates the rotational effect small variations in the DM can have on the autocorrelation of FRB waterfalls.}\label{fig:Angle-at-different}
\end{figure}
\end{center}



\subsection{\texorpdfstring{Unresolved bursts and slope vs. drift rate ambiguity}{}}\label{sec:unresolvedfrbs}

In practice it is difficult to be certain whether an FRB is resolved in time or not, and in several cases the difference between sub-burst slope and sub-burst drift measurements is ambiguous. For example, the waterfall of an unresolved pulse train will go through our analysis and produce a sub-burst slope measurement, while it is, in fact, a sub-burst drift rate measurement. While, as mentioned earlier, we make every effort to separate FRBs into distinct sub-bursts, it will be impossible to do this with bursts that may have originally been composed of multiple components but appear completely unresolved in the data. Here, we discuss ways through which an FRB becomes unresolved and the connection between the sub-burst slope and sub-burst drift rate measurements within the context of our model.

An FRB that was emitted as distinct components in time can appear unresolved in data either through a limited time resolution (e.g., Fig. 1 of both \citealt{Michilli2021} and \citealt{Bhardwaj2021}), intrachannel smearing from incoherent dedispersion, limited S/N (e.g., Fig. 7 of \citealt{Gourdji2019}), and the blending of components through scatter broadening (e.g., Sec. 4.3 of \citealt{Day2020}). In the sample of bursts used for this analysis, bursts numbered 13 and 15 in Figure \ref{fig:frb121102card}, and bursts 14, 18, and 23 in Figure \ref{fig:frb180916card} are suggestive of being composed of multiple sub-bursts, and others are likely indistinguishable from multi-bursts due to their time resolution, just as in the aforementioned example from \citealt{Gourdji2019}.

This raises the question of how a majority of the sub-burst slopes shown in Figure \ref{fig:sub-burst_drift} agree with the expected relation when more outliers might be expected. Firstly, it is possible that the presence of bursts exhibiting large variations in their measurements is an indication of outlying FRBs that were incorrectly measured. Another possibility is that as the sub-bursts are smeared together, the measured duration will necessarily increase, and, since sub-bursts are mostly observed to drift downwards in frequency, the slope of the resulting autocorrelation will also be shallower. While these scenarios qualitatively describe what we should expect when measuring pulse trains as a single sub-burst, we note that our dynamical model predicts that groups of sub-bursts emitted close to each other in time will follow the same relationship as individual sub-bursts when the difference in delay time is small. This follows from equation (6) of \citealt{Rajabi2020b} that relates different time and frequency intervals between the FRB and observer reference frames
\begin{equation}
\Delta t_{\text{D}}=-t_{\text{D}}\Big(\frac{\Delta\nu_{\text{obs}}}{\nu_{\text{obs}}}-\frac{\Delta\tau'_{\text{D}}}{\tau'_{\text{D}}}\Big)\label{eq:timeints},
\end{equation}
where as before $t_{\text{D}}$ and $\tau'_{\text{D}}$ are the delay time between the trigger signal and the appearance of a sub-burst as measured in the observer's and FRB frames, respectively, and $\nu_\text{obs}$ is the observed frequency of the burst. The quantities $\Delta \nu_\text{obs}$, $\Delta t_\text{D}$, and $\Delta \tau'_\text{D}$ account for variations in these parameters. The relation central to this paper shown in equation \ref{eq:drift} is obtained when $\Delta \tau'_\text{D} \simeq 0$, a condition that applies to groups of sub-bursts that occur closely in time as well as individual sub-bursts.

Given the possible agreement between the trends of the sub-burst slope and sub-burst drift measurements for the FRBs we have considered, we can use the following physical interpretation in the context of our model. When observing a resolved sub-burst signal, the frequency bandwidth of the signal is determined by the dynamical velocity of the emitting material. When multiple sub-bursts are observed to occur close to each other in time (such as within a single FRB event), the drift rate of the pulse train depends on the delay time between the FRB trigger and the FRB event. When variations in the delay time are small the sub-burst drift rate will follow the same trend as the slope measurements of the individual sub-bursts. In this case, if the sub-bursts are unresolved then it will be difficult to distinguish a slope measurement from a drift rate measurement. The lack of a significant number of outliers in Figure \ref{fig:sub-burst_drift} suggests that the situation with small variations in delay time is common.

\subsection{\texorpdfstring{DM variations as a rotation of the autocorrelation function}{}}\label{sec:dmrotates}
Following the study of the variation of measurements over ranges of plausible DMs discussed in Sec. \ref{sec:effectofdm}, different DM choices can be modeled as rotations of the autocorrelation function of the burst. As an example, we show in Figure \ref{fig:rotation_stamp} two bursts each at two choices of DM. In a given waterfall we see that the shape of the burst can `distort' due to the $\nu^{-2}_\mathrm{obs}$ dependence on the dispersion, while that of the autocorrelation remains practically the same except for experiencing a rotation. To characterize this further we consider the sub-burst angle parameter (as opposed to the corresponding sub-burst slope derived from said angle) defining the orientation of the fitted ellipsoid's semi-major axis measured counterclockwise from the positive frequency axis of the autocorrelation function (see Appendix \ref{app:autocorr}). The sub-burst angles from FRB~20180916B are plotted against the corresponding temporal durations derived from the underlying two-dimensional Gaussian fits in Figure \ref{fig:Angle-at-different}. This shows that across different DMs the measured duration varies little while the angle is offset by a constant level from values at other DMs.
We can demonstrate this using equations (\ref{eq:drift}) and (\ref{eq:drifttheta_SM}) to find that the slope angle $\theta$ is related to the sub-burst duration through 
\begin{equation}
\theta = \arctan\left(\frac{1}{A}\frac{\nu_\mathrm{res}}{\nu_\mathrm{obs}}\frac{t_{\mathrm{w}}}{t_\mathrm{res}}\right),\label{eq:theta_SM}
\end{equation}
where as before $A\equiv\tau_\mathrm{w}^\prime/\tau_\mathrm{D}^\prime$, and $\nu_\mathrm{res}$ and $t_\mathrm{res}$ are the frequency and time resolutions of the waterfall. We also approximated $\nu_\mathrm{obs}$ to be constant, which is adequate for this purpose. We find that the chosen fit obtained with equation (\ref{eq:theta_SM}) for the sub-bursts at $\Delta\mathrm{DM}=0$ (i.e., the solid curve in Figure \ref{fig:Angle-at-different}) is also satisfactory for angles corresponding to the different $\Delta\mathrm{DM}$ values when a simple offset angle (i.e. a rotation) is applied. Similar trends appear to hold for the other two sources considered, however it is most clear in the example of FRB~20180916B.

Measuring the sub-burst angle instead of the slope during analysis avoids the discontinuity of slope measurements around $\theta=0$ or $\pi$, where its magnitude approaches infinity, however the uncertainty of the physically relevant quantity will still be large. The behaviour of slope measurements derived from the parameter angle in the context of autocorrelation noise is discussed in more detail in \citet{PleunisThesis2020} as well as in the \texttt{dfdt}\footnote{\url{https://github.com/zpleunis/dfdt}} package.


\subsection{Uncertainty due to frequency band masking}

In addition to signal noise, the waterfall analyses were complicated by missing frequency bands of data, which would sometimes overlap with the frequency extent of the sub-burst under consideration. In this section we assess the extent of the uncertainty introduced by the missing frequency band data by (1) artificially masking (zero-padding) various trial Gaussian signals of known orientations and characteristic widths, (2) processing them through our pipeline, and (3) comparing the extracted sub-burst slope and duration parameters to the generating parameters.

Consider for example Burst 23 of FRB~20180916B \citep{CHIME2020a} pictured, along with its two-dimensional autocorrelation, in Figure \ref{fig:frb180916card}. Three frequency bands of data are absent from the original data in this burst, and the total missing bandwidth (as a fraction of the frequency extent of the sub-burst) is higher than the fractional bandwidth typically absent from sub-bursts analyzed in the paper.

\begin{centering}
    \begin{figure}\hspace*{-0.5cm}
        \centering
        \includegraphics[width=1.\columnwidth]{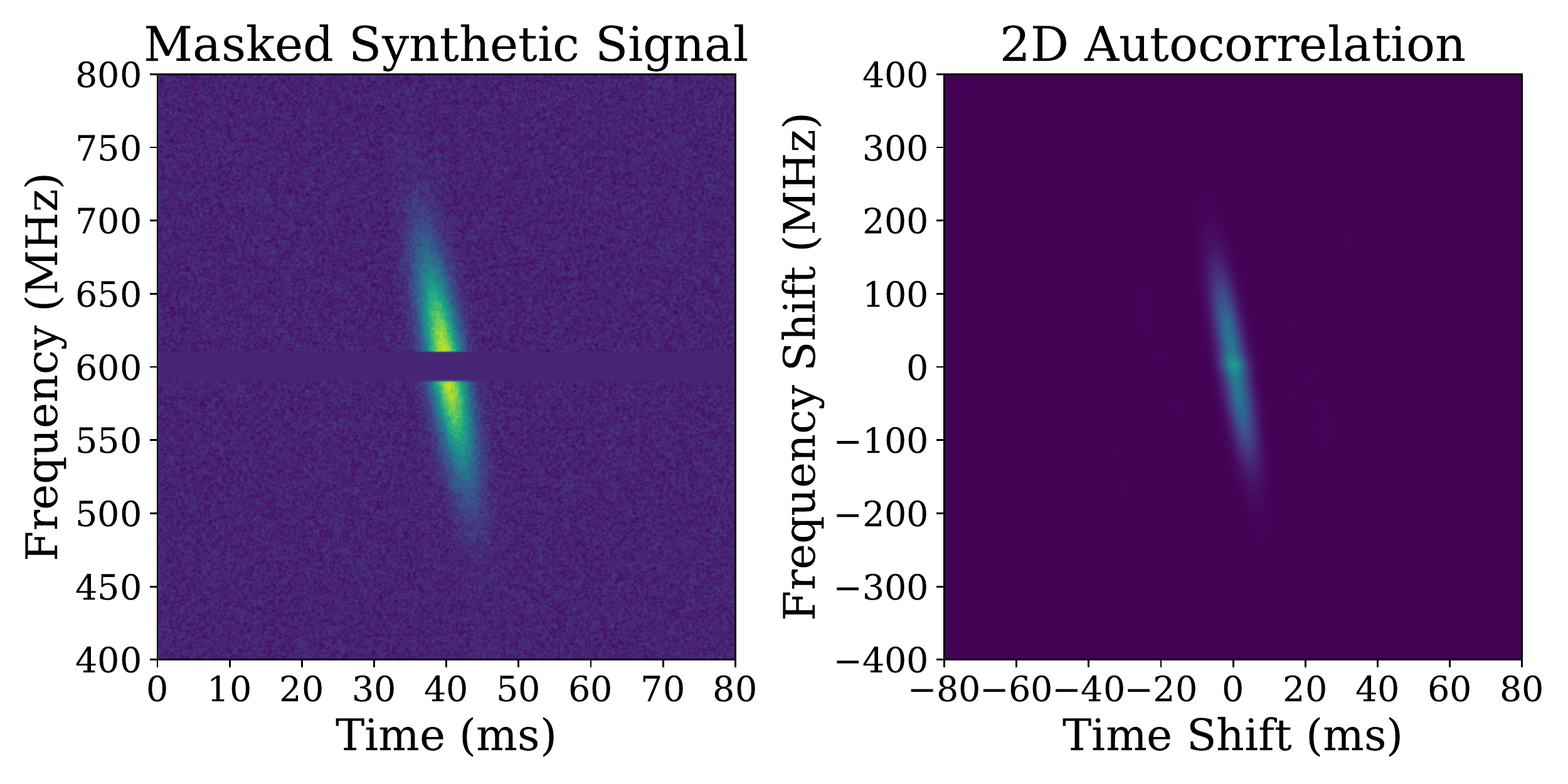}
        \caption{Synthetic Gaussian signal with a masked (zeroed) band (left), and 2D autocorrelation of masked signal (right). The signal shown approximately matches the characteristics of Burst 23 depicted in Figure \ref{fig:frb180916card} in each of their temporal widths, their frequency extents, their inclinations, and their total missing frequency bandwidths.}
        \label{fig:PipelineVisual}
    \end{figure}
\end{centering}

To estimate the effect of missing frequency channels on our analysis we construct an artificial burst similar to Burst 23 from a Gaussian with a standard deviation along the semi-major axis of $a=67 \text{ MHz}$ (90 pixels), standard deviation along the semi-minor axis $b=2.2 \text{ ms}$ (15 pixels), $\theta=10^{\circ}$ (inclination from vertical), and with stochastic noise of amplitude 25\% that of the Gaussian amplitude. We perform our analysis on a 2D array with dimensions $540\times540$ pixels, having horizontal and vertical resolutions of 6.75 px/ms and 1.35 px/MHz, respectively. As a first test, we mask a band of width $18.5 \text{ MHz}$ (25 pixels) through the center of the burst and pass this zero-padded signal through our pipeline. 

The fitting procedure on the 2D autocorrelation returns $a_{\mathrm{fit}}=102.7$ pixels, $b_{\mathrm{fit}}=14.8$ pixels, and $\theta_{\text{fit}}=9.83^{\circ}$. The process is visualized in Figure \ref{fig:PipelineVisual}. For such a small inclination angle, the percentage error in $t_{\mathrm{w}}$ is very close to that of $b$, and is (in this case) approximately 1\%. The corresponding percentage error in the sub-burst slope $\mathrm{d}\nu_{\mathrm{obs}}/\mathrm{d}t_\mathrm{D}$ is 1.7\%.

We can generalize this test by shifting the frequency masking band of Figure \ref{fig:PipelineVisual} vertically. Upon doing so, we find that the error is independent of the frequency band’s vertical position. The percentage error for the burst duration is found to be $\simeq\left(-1.4\pm0.4\right)\%$, where the $\pm0.4\%$ uncertainty applies to all band vertical positions tested, while the corresponding error in the angle is $\simeq\left(-1.1\pm0.7\right)\%$.

If we rotate the burst of Figure \ref{fig:PipelineVisual}, while retaining the central band mask of $18.5 \text{ MHz}$ (25 pixels) on burst centre, we observe a linear enhancement of error with increasing orientation. The effect is, however, a negligible one: for every burst rotation by $10^{\circ}$, the duration error increases by only $0.45\%$, while the orientation angle error decreases (or increases in magnitude) by only $0.12^{\circ}$. At a $30.0^{\circ}$ burst angle, the sub-burst slope error is only $4\%$.

\subsection{The narrow-band nature of the emission process}\label{sec:narrownature}

\begin{center}
    \begin{figure}
        \centering
        \hspace*{-0.5cm}
        \includegraphics[trim=0 0 0 0, clip,width=1.\columnwidth]{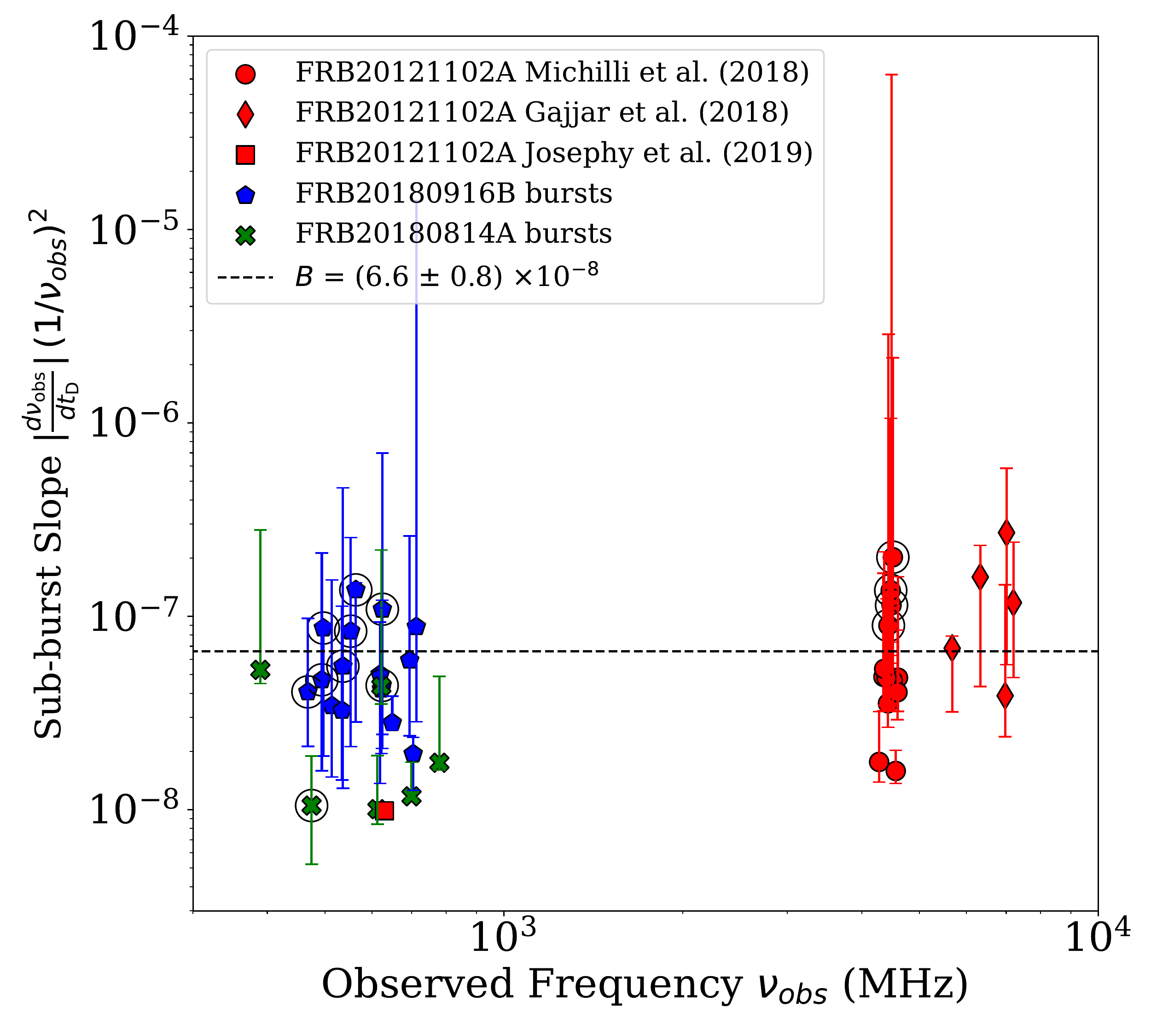}
        \caption{A plot of $\left|d\nu_\mathrm{obs}/dt_\mathrm{D}\right|\left(1/\nu_\mathrm{obs}^2\right)$ vs. $\nu_\mathrm{obs}$ for FRB~20121102A (red circles, diamonds and square from \citealt{Gajjar2018}, \citealt{Michilli2018} and \citealt{CHIME2019c}, respectively), FRB~20180916B (blue pentagons; \citealt{CHIME2020a}) and FRB~20180814A (green crosses; \citealt{CHIME2019a}). The broken black line is for a fit to a constant $B$ on the combined data for the three sources, with $B\equiv\left(\tau_\mathrm{D}^\prime\nu_0\right)^{-1}=\left(6.6\pm0.8\right)\times10^{-8}$. The capped lines at each point represent the range of possible values due to the range of sub-burst slopes measured at different DMs, as discussed in Section \ref{sec:effectofdm}.} 
        \label{fig:sub-burst_drift_SM}
    \end{figure}
\end{center}

As discussed in Section \ref{sec:model}, results from \citet{Gajjar2018} and \citet{Hessels2019} for FRB~20121102A point to a rest frame frequency of emission $\nu_0$ that does not change significantly from burst to burst. The results presented here can be shown to also be consistent with a narrow-band emission process by inserting equation (\ref{eq:tw_SM}) into equation (\ref{eq:drift}) to obtain
\begin{align}
    \frac{1}{\nu_\mathrm{obs}^2}\frac{d\nu_\mathrm{obs}}{dt_\mathrm{D}} = -\frac{1}{\nu_0\tau_\mathrm{D}^\prime},\label{eq:drift_SM}
\end{align}
\noindent for the sub-burst slope (normalized to $\nu_\mathrm{obs}^2$), which is then predicted to be independent of $\nu_\mathrm{obs}$ and scale inversely with $\nu_0$. Figure \ref{fig:sub-burst_drift_SM} shows the corresponding plot using the same data as in Figure \ref{fig:sub-burst_drift}. The broken black line is for a fit to a constant $B$ on the combined data for the three sources, with $B\equiv\left(\tau_\mathrm{D}^\prime\nu_0\right)^{-1}=\left(6.6\pm0.8\right)\times10^{-8}$. While there is some scatter in the data, the result is consistent with the expected lack of dependency on $\nu_\mathrm{obs}$. Any deviation could easily be accounted for with the uncertainty on the DMs and inherent variations in $\tau_\mathrm{D}^\prime$. The combination of this result with the temporal narrowing and sad trombone effects discussed in Section \ref{sec:model} for FRB~20121102A provides evidence for the narrow-band nature of the emission process. 

\section{Conclusion}

We demonstrate a method of studying the sub-burst slope in the context of DM variations from burst to burst and over time by adopting large ranges of possible DMs when measuring spectro-temporal properties of FRBs. This method reveals that even given a wide range of possible DMs for each burst from an FRB source, the slope of an individual sub-burst is inversely proportional to its temporal duration. Furthermore, for the three sources considered in this work, namely FRB~20121102A, FRB~20180916B and FRB~20180814A, significant overlap between the inverse trends found is consistent with the three relationships having a nearly identical scaling. That is, the same law can be used to describe sub-bursts from all three sources, though careful analyses over larger data sets at different frequencies would be needed to verify this. Additionally, this result suggests that the sub-burst slope law may be a useful tool for simplifying studies that require large samples of FRBs, by providing a single small adjustment DM to dedisperse waterfalls to without the complexity of verifying each burst's DM.

We believe that the simplest explanation for the existence of this trend is that the emission mechanism of these FRB sources is narrow-band in nature, which would be consistent with earlier models based on Dicke's superradiance \citep{Houde2018b,Rajabi2019,Rajabi2020}. Such a mechanism requires a trigger, which leaves room for magnetar-centric models of FRBs. To further study the relationship between the sub-burst slope and temporal duration future analyses of FRBs from all known repeater sources can be performed in the manner presented here. A large sample of sources helps to constrain the uncertainties due to variations in DM, and necessitates convenient and public access to FRB data. 

Finally, we note that our discovery of a shared sub-burst slope law among these three sources suggests that this could be a universal property among repeating FRBs or at least a significant subclass of them. If deviations from this relationship exist, then it is likely the sub-burst slope law can serve as a classification tool for FRBs by discriminating sources that follow this law from those that do not. This not only motivates further searches but also provides a new tool to study and categorize FRBs based on their underlying physical mechanism.


\begin{figure*}
\centering
\includegraphics[width=0.9\textwidth]{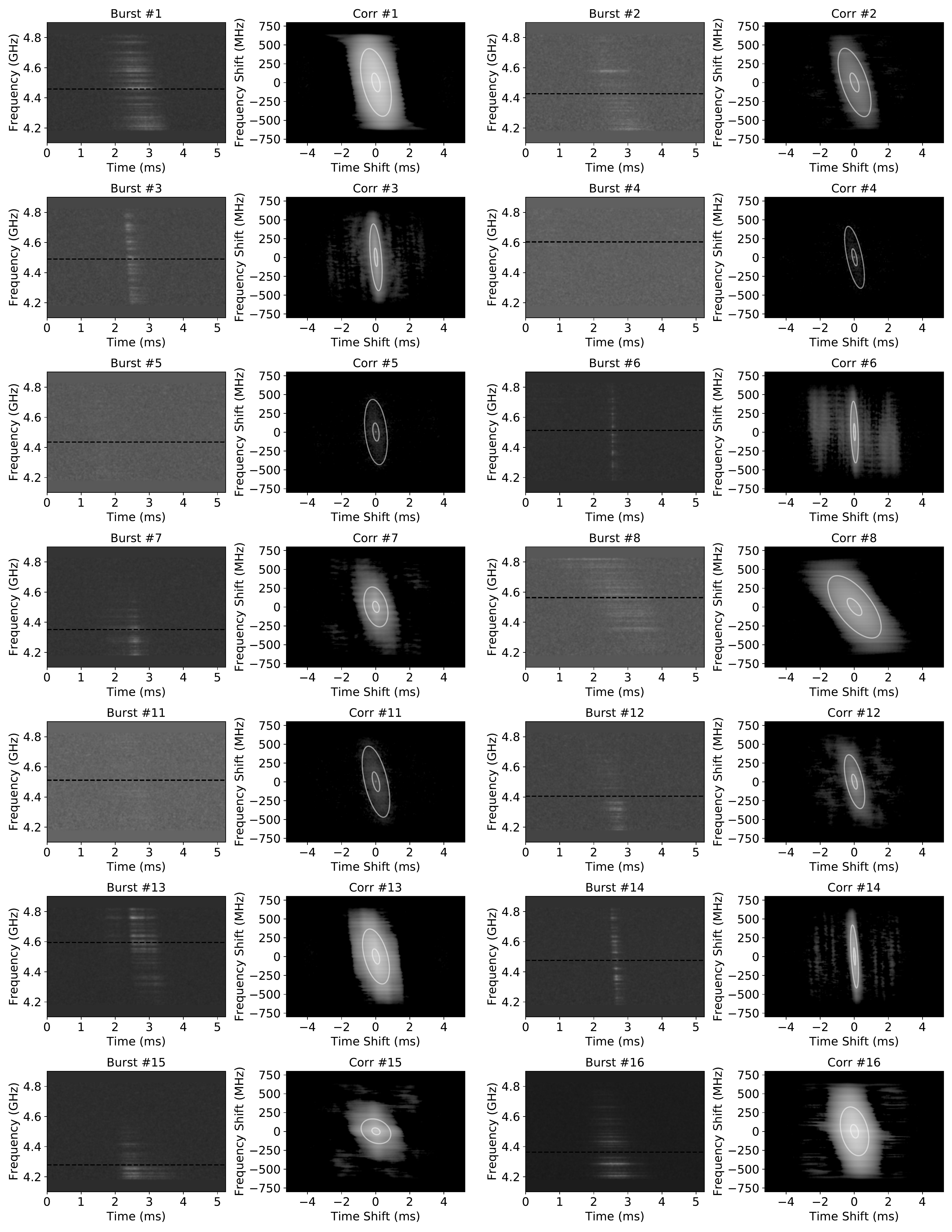}
\caption{Dynamic spectra (first and third columns) and corresponding autocorrelation functions (second and fourth columns) for FRB~20121102A bursts at a frequency of approximately 4--5~GHz from \citet{Michilli2018}. The dynamic spectra were dedispersed with a $\mathrm{DM}=559.7\;\mathrm{pc\,cm}^{-3}$ and the dashed horizontal line in the waterfall denotes the center frequency $\nu_\mathrm{obs}$ used for the analysis. The autocorrelation functions are modelled with a 2D Gaussian ellipsoid whose one- and two-standard deviation levels are shown using the white contours.}\label{fig:frb121102card}
\end{figure*}


\begin{figure*}
\centering
\includegraphics[width=0.65\textwidth]{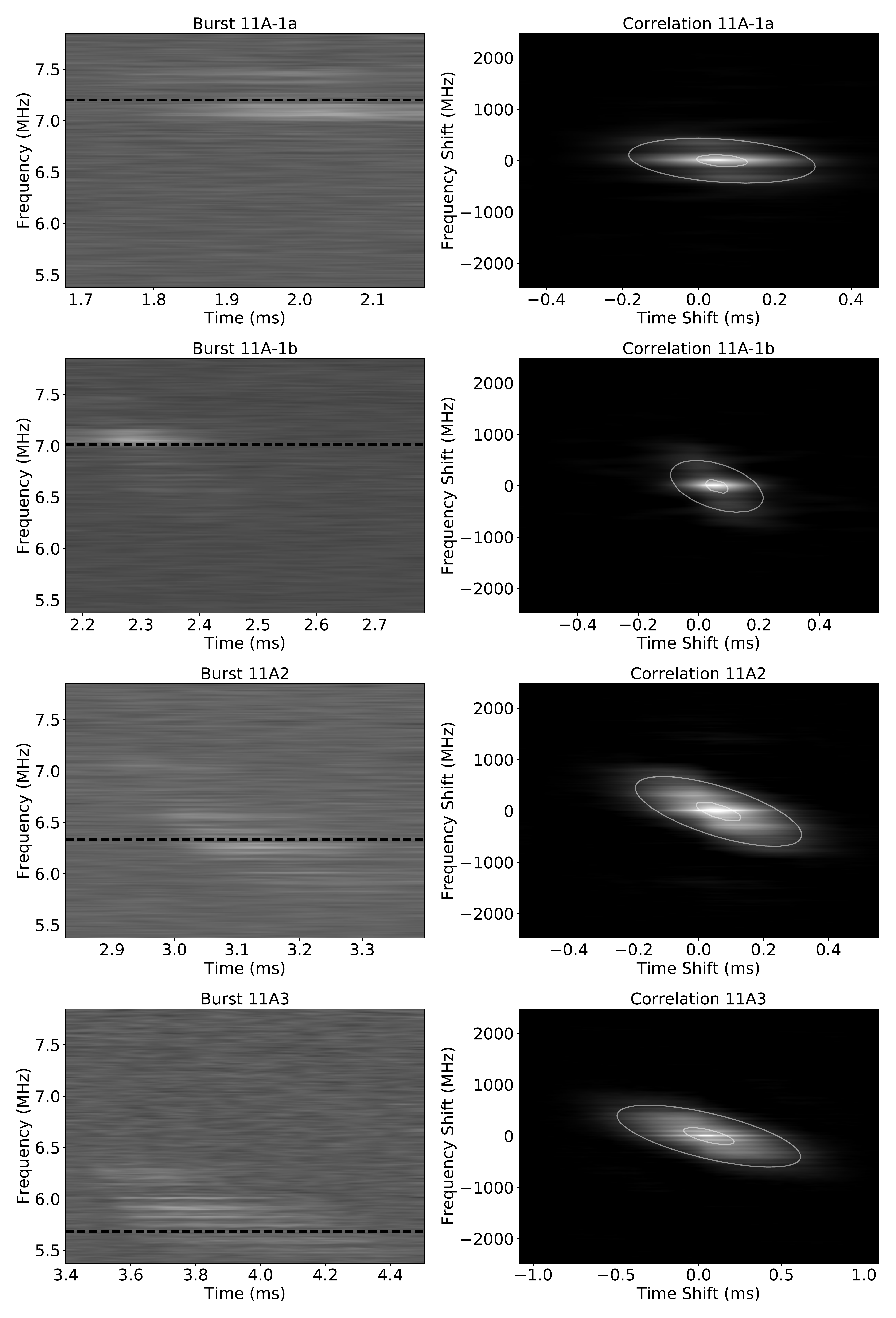}
\includegraphics[width=0.65\textwidth]{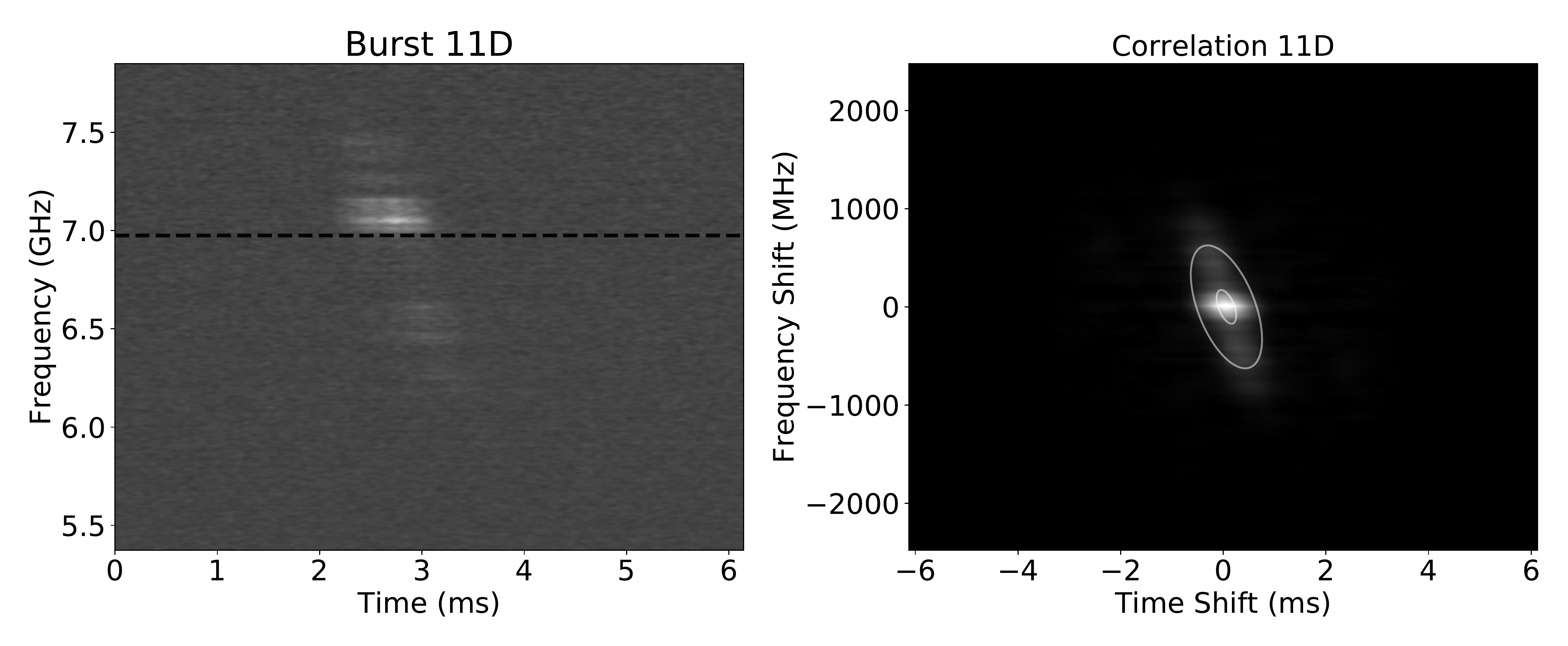}
\caption{Same as Figure \ref{fig:frb121102card} but for the FRB~20121102A data at approximately 5--8~GHz published in \citet{Gajjar2018} and dedispersed with a $\mathrm{DM}=565\;\mathrm{pc\,cm}^{-3}$. The top four sub-bursts are taken from one event, i.e., Burst 11A. Note that the time axes for the autocorrelation functions do not all share the same range, which distorts their relative appearance.}\label{fig:frb121102card_2}
\end{figure*}

\begin{figure*}
\centering
\includegraphics[width=0.95\textwidth]{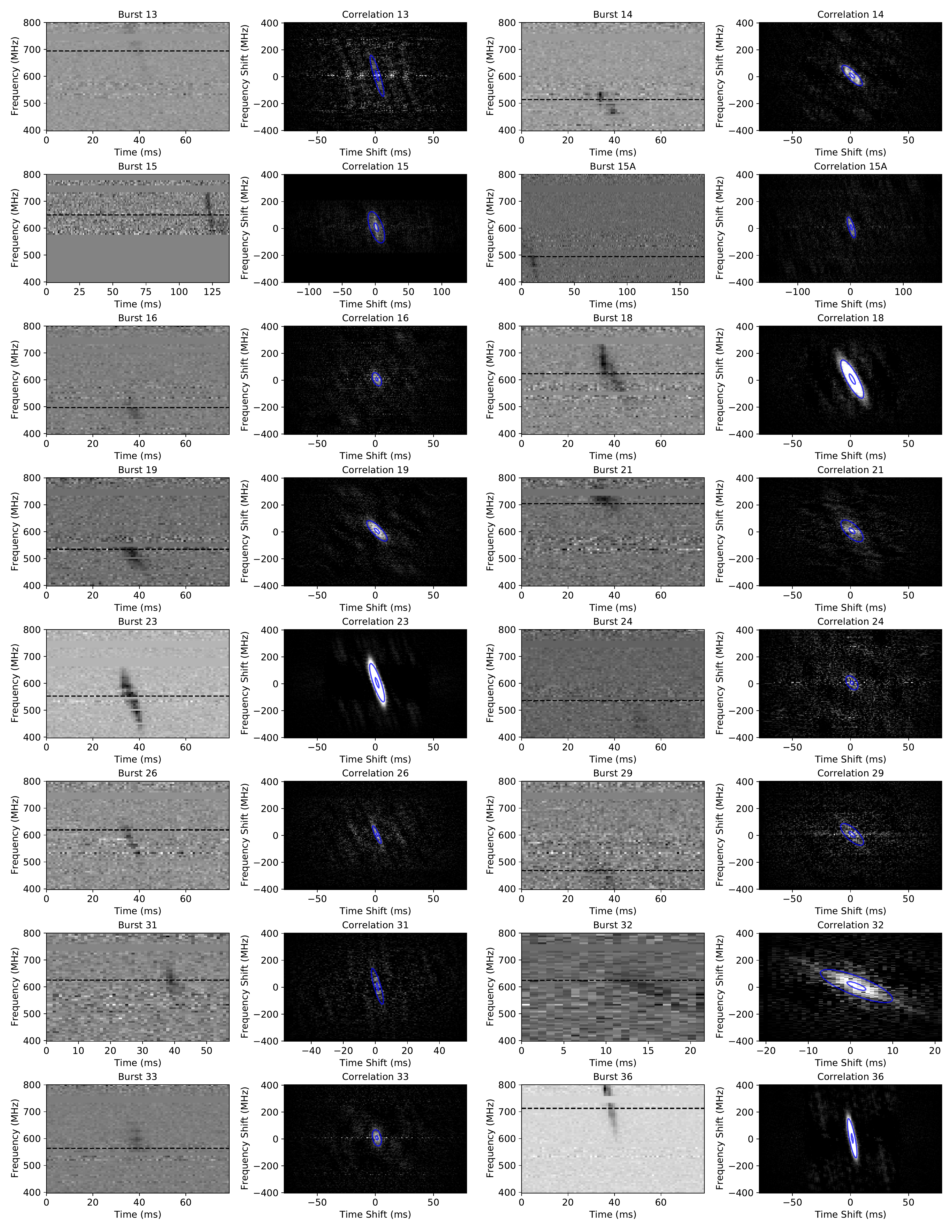}
\caption{Same as Figure \ref{fig:frb121102card} but for FRB~20180916B taken from \citet{CHIME2020a}. These data were dedispersed with a $\mathrm{DM}=348.82\;\mathrm{pc\,cm}^{-3}$. Note that the time axes for the autocorrelation functions do not all share the same range, which distorts their relative appearance.}\label{fig:frb180916card}
\end{figure*}

\begin{figure*}
\centering
\includegraphics[width=0.8\textwidth]{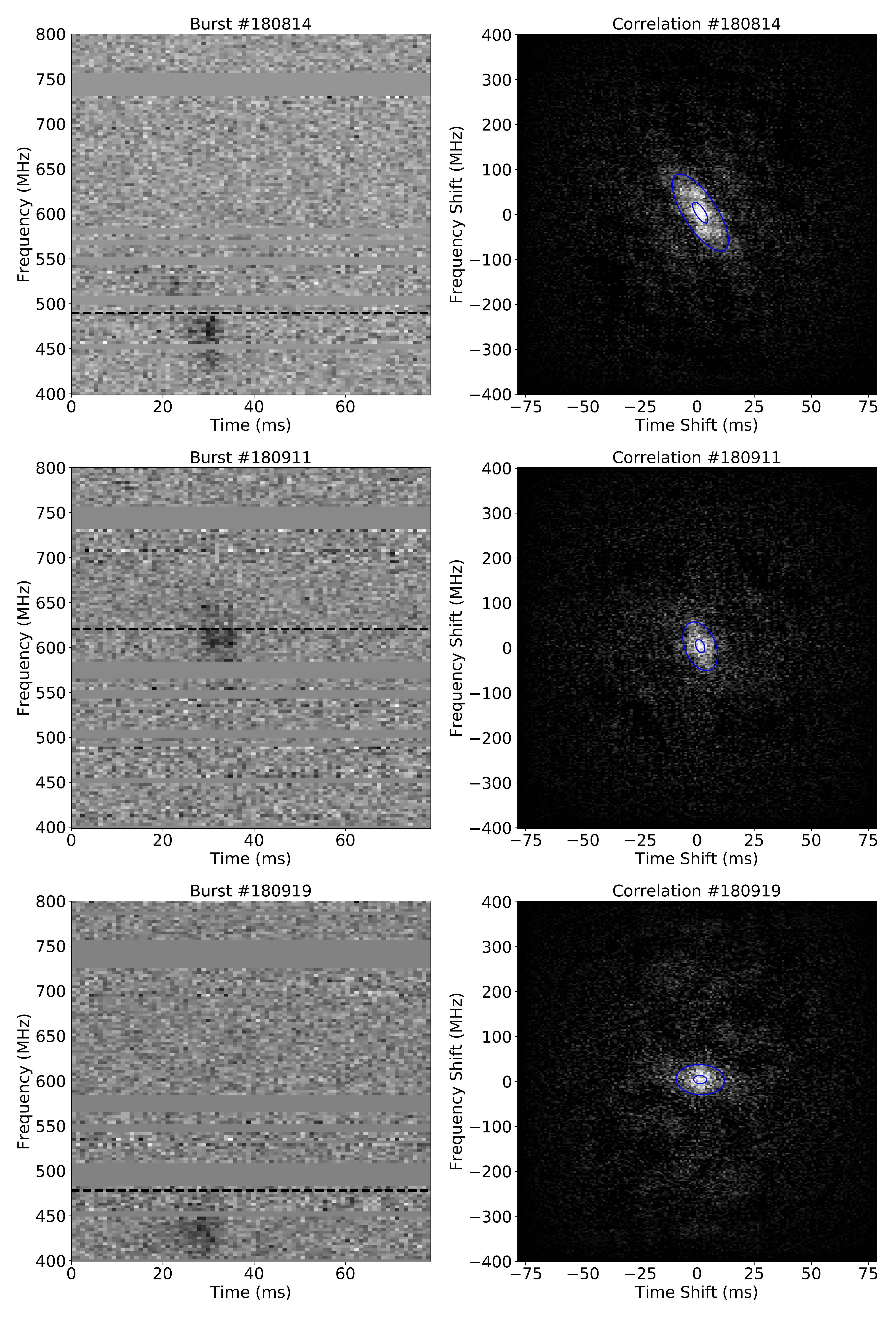}
\caption{Same as Figure \ref{fig:frb121102card} but for FRB~20180814A taken from \citet{CHIME2019a}. These data were dedispersed with a $\mathrm{DM}=188.9\;\mathrm{pc\,cm}^{-3}$.}\label{fig:frb180814card_1}
\end{figure*}

\begin{figure*}
\centering
\includegraphics[width=0.8\textwidth]{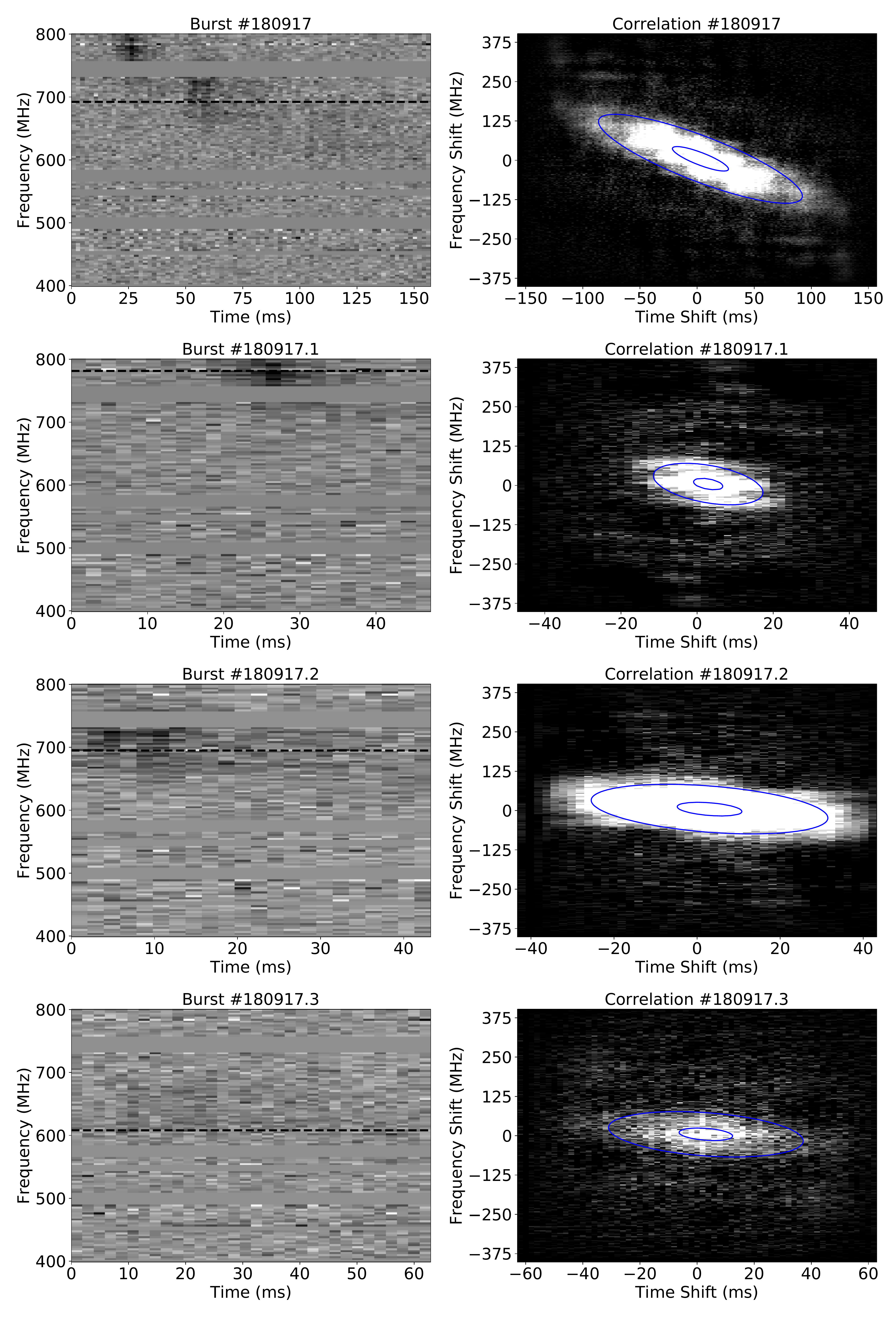}
\caption{Same as Figure \ref{fig:frb180814card_1} but for Burst \#180917 of FRB~20180814A taken from \citet{CHIME2019a}. The whole event is shown on the top row (not used for Figure 1 of main text), while its three separate sub-bursts are detailed in the bottom three (all used for Figure 1 of main text). Note that the time axes for the autocorrelation functions do not all share the same range, which distorts their relative appearance.}\label{fig:frb180814card_2}
\end{figure*}

\section*{Acknowledgements}
The authors are grateful to Z. Pleunis and S. Tendulkar from the CHIME\-/FRB Collaboration for their help in accessing and analyzing the data for FRB~20180916B and FRB~20180814A. The authors are also grateful to the referees for the extensive feedback that significantly shaped the final manuscript. M.H. is grateful for the hospitality of Perimeter Institute where part of this work was carried out. M.H.'s research is funded through the Natural Sciences and Engineering Research Council of Canada Discovery Grant RGPIN-2016-04460. F.R.'s research at Perimeter Institute is supported in part by the Government of Canada through the Department of Innovation, Science and Economic Development Canada and by the Province of Ontario through the Ministry of Economic Development, Job Creation and Trade. F.R. is in part financially supported by the Institute for Quantum Computing. C.M.W. and A.M. are supported by the Natural Sciences and Engineering Research Council of Canada (NSERC) through the doctoral postgraduate scholarship (PGS D).

\section*{Data Availability}

The data pipeline is made available at \url{https://github.com/mef51/subdriftlaw} and maintained by M.A.C. Aggregate data of the bursts and the code for the figures are also available. Data of the FRB spectra are available either publicly or via the authors of their respective publications. The figures in this paper were prepared using the {\tt matplotlib} package \citep{Hunter2007}.




\bibliographystyle{mnras}
\bibliography{scibib}



\appendix

\section{Autocorrelation Analysis}\label{app:autocorr}

We discuss here the process of preparing and obtaining measurements from the dynamic spectra of bursts, based on the autocorrelation technique described in \citet{Hessels2019}. 

As mentioned in Section \ref{sec:effectofdm} when a waterfall consists of a train of multiple sub-bursts we separate the components and measure the slope and temporal duration of each sub-burst separately. The dynamic spectra of every sub-burst used in this analysis with its autocorrelation is shown in Figures \ref{fig:frb121102card} -- \ref{fig:frb180814card_2}.

The pipeline that every sub-burst undergoes is written in Python and consists of computing the autocorrelation of the signal, fitting a two-dimensional (2D) Gaussian to the resulting autocorrelation function, and a calculation of the physical quantities of interest from the fit parameters: namely the sub-burst slope and temporal duration. The autocorrelation of the waterfall measures the self-similarity of the sub-burst in frequency-time space and for FRBs can be approximated by an ellipsoid with an intensity that follows a 2D Gaussian \citep{Hessels2019}. Before computing the autocorrelation and depending on the source and/or burst, some noise removal is performed. For the bursts from FRB~20121102A and FRB~20180916B this is done by subtracting from the entire spectrum a background signal obtained from a time-average of twenty or so samples taken prior to the burst. For FRB~20180814A, due to the raw format the bursts are provided in, a noise mask was acquired through correspondence with members of the CHIME/FRB Collaboration and the channels are normalized by the standard deviation of the intensity. Missing or blocked out frequency channels in dynamic spectra (e.g., because of radio frequency interference (RFI)) are zeroed out before performing the autocorrelation. 

The computation of the autocorrelation function is facilitated and sped up by using a Fast Fourier Transform (FFT) of the waterfall, which is then squared and inverted (through an FFT) back to the frequency-time domain \citep{Numerical2007}. The autocorrelation function is then modelled with the following functional form for a rotated 2D Gaussian
\begin{align}
    G\left(x,y\right) & = C\exp\left\{-\frac{1}{2}\left[x^2\left(\frac{\cos^{2}\theta}{b^2}+\frac{\sin^{2}\theta}{a^2}\right)\right.\right.\nonumber\\
    & \left.\rule{0mm}{6mm}\left.+2xy\sin\theta\cos\theta\left(\frac{1}{b^2}-\frac{1}{a^2}\right)+y^2\left(\frac{\sin^{2}\theta}{b^2}+\frac{\cos^{2}\theta}{a^2}\right)\right]\right\},\label{eq:Gaussian_SM}
\end{align}

\noindent with the free parameters $C$, $a$, $b$, and $\theta$ for, respectively, the amplitude, the semi-major and semi-minor axes (i.e., the standard deviations) of the ellipsoid, and the sub-burst angle for the orientation of the semi-major axis measured counterclockwise from the positive $y$-axis. More precisely, the $x$- (i.e., for the time lag) and $y$-axes (i.e., for the frequency lag) are respectively oriented horizontally and vertically on the autocorrelation plots shown in Figures \ref{fig:frb121102card}--\ref{fig:frb180814card_2}. To find these parameters we use the\texttt{ scipy.optimize.curve\_fit }package, which performs a non-linear least squares fit. The package also returns a covariance matrix, which is used to calculate the uncertainty of the fitted parameters. These uncertainties are then scaled by the square-root of the reduced-$\chi^{2}$ computed from the residual between the autocorrelation function and its Gaussian fit. We note again that the uncertainty calculated this way does not capture nearly the entire error budget which depends more significantly on the error in the DM (discussed in Section \ref{sec:effectofdm}) as well the parts of the burst spectra that have been masked out and the shape of its autocorrelation. 

Using the angle $\theta$, the sub-burst slope is calculated via
\begin{equation}
\frac{d\nu_{\mathrm{obs}}}{dt_\mathrm{D}} = -\frac{\nu_{\mathrm{res}}}{t_{\mathrm{res}}}\cot\theta,\label{eq:drifttheta_SM}
\end{equation}
where $\nu_{\mathrm{res}}$ and $t_{\mathrm{res}}$ are the frequency and time resolutions of the waterfall. We obtain the sub-burst duration from the fit parameters through
\begin{equation}
t_{\mathrm{w}}=t_{\mathrm{res}}\frac{ab}{\sqrt{b^{2}\sin^{2}\theta+a^{2}\cos^{2}\theta}}.\label{eq:t_w_SM}
\end{equation}
These expressions are also used to propagate the fit parameter uncertainties to the values of $d\nu_{\mathrm{obs}}/dt_\mathrm{D}$ and $t_{\mathrm{w}}$. These uncertainties are used to confirm the assertion that DM variations are the largest source of error, as stated in Section \ref{sec:effectofdm}.

The observation frequency $\nu_{\mathrm{obs}}$ of each sub-burst is estimated via an intensity-weighted average of the spectrum over the whole time range. While this decreases the accuracy of the estimate as opposed to using just the on-pulse region, we find it has little bearing on the result. To fit equation (\ref{eq:drift}) we used the \texttt{scipy.odr.RealData} package, which uses orthogonal distance regression and the uncertainties on the data to find a fit. 

\section{\texorpdfstring{Determination of $\beta^+$, $\nu_0$ and $\Delta\beta^\prime$}{}}\label{sec:sourceparams}

The equations presented in this section apply to cases where the source of radiation travels directly toward or away from the observer. 

For the determination of the maximum speed of an FRB rest frame toward the observer $\beta^+>0$ and $\nu_0$, the frequency of emission within it, we can generally set $\beta^-=-a\beta^+$ with $a\ge0$ for the greatest (i.e., most negative) speed away from the observer. Using the relativistic Doppler shift formula \citep{Rybicki1979} for the corresponding frequencies in the observer's rest frame
\begin{align}
    \nu_\mathrm{obs}^\pm = \nu_0\sqrt{\frac{1+\beta^\pm}{1-\beta^\pm}},\label{eq:Doppler_SM}
\end{align} 
\noindent we find that
\begin{align}
    \beta^+ & = \left(\frac{1+a}{2a}\right)\left(\frac{{\nu_\mathrm{obs}^+}^2+{\nu_\mathrm{obs}^-}^2}{{\nu_\mathrm{obs}^+}^2-{\nu_\mathrm{obs}^-}^2}\right)\nonumber \\
    & \quad\times\left[1-\sqrt{1-\frac{4a}{\left(1+a\right)^2}\left(\frac{{\nu_\mathrm{obs}^+}^2-{\nu_\mathrm{obs}^-}^2}{{\nu_\mathrm{obs}^+}^2+{\nu_\mathrm{obs}^-}^2}\right)^2}\right]\label{eq:beta+_SM}\\
    \nu_0^2 & = \nu_\mathrm{obs}^+\nu_\mathrm{obs}^-\sqrt{\frac{1-\left(1-a\right)\beta^+ -a{\beta^+}^2}{1+\left(1-a\right)\beta^+ -a{\beta^+}^2}}.\label{eq:nu0_SM}
\end{align}
The discussion in Section \ref{sec:results} where the FRB rest frames span the range $\pm\beta^+$ corresponds to the case $a=1$, which reduces equations (\ref{eq:beta+_SM})-(\ref{eq:nu0_SM}) to equations (\ref{eq:beta+})-(\ref{eq:nu0}) of the main text. 

For the determination of $\Delta\beta^\prime$, we start by considering that for a signal initially observed at frequency $\nu_\mathrm{obs}$ a velocity change $\Delta\beta$ in the observer's rest frame will be accompanied by a change $\delta\nu_\mathrm{obs}$ in frequency given by
\begin{align}
    \frac{\delta\nu_\mathrm{obs}}{\nu_\mathrm{obs}} = \frac{\Delta\beta}{1-\beta^2},\label{eq:dbeta_SM}
\end{align}
\noindent where $\beta$ is the initial velocity relative to the observer. Using the special relativistic velocity addition law \citep{Rybicki1979} we can relate the velocity changes in the observer and FRB rest frames through
\begin{align}
    \Delta\beta = \Delta\beta^\prime\left(\frac{1-\beta^2}{1+\beta\Delta\beta^\prime}\right),\label{eq:dbeta'_SM}
\end{align}
\noindent with $\Delta\beta^\prime$ the corresponding velocity change in the FRB frame.

Allowing for the motions within the FRB rest frame to span the range $\pm\Delta\beta^\prime$ (with $\Delta\beta^\prime\ge0$; for simplicity, we assume a symmetric velocity range about zero), while using equations (\ref{eq:Doppler_SM}) (to express $\beta$ as a function of $\nu_\mathrm{obs}$ and $\nu_0$) and (\ref{eq:dbeta_SM})-(\ref{eq:dbeta'_SM}), we find the following relation for the total observed bandwidth covered by the corresponding signals
\begin{align}
    \frac{\Delta\nu_\mathrm{obs}}{\nu_\mathrm{obs}} = 2\Delta\beta^\prime\left[1-{\Delta\beta^\prime}^2\left(\frac{\nu_\mathrm{obs}^2-\nu_0^2}{\nu_\mathrm{obs}^2+\nu_0^2}\right)^2\right]^{-1}.\label{eq:dfreq_SM}
\end{align}
\noindent Equation (\ref{eq:spectral_width}) follows from this relation, which reaches a maximum value when $\nu_\mathrm{obs}=0$ or $\nu_\mathrm{obs}\gg\nu_0$. While equation (\ref{eq:dfreq_SM}) shows little variations whenever $\Delta\beta^\prime\ll1$, it could, in principle, be used to evaluate the FRB rest frame frequency $\nu_0$ independently of equation (\ref{eq:nu0_SM}) since it reaches a minimum of $2\Delta\beta^\prime$ at $\nu_\mathrm{obs}=\nu_0$. However, the effect is probably too small (on the order of $1\%$ for FRB~20121102A) to be measurable given the scatter inherent to FRB data. 

\bsp	
\label{lastpage}
\end{document}